\documentclass[floats,floatfix,showpacs,amssymb,prd,twocolumn,superscriptaddress,nofootinbib,nolongbibliography,reprint]{revtex4-1}

\usepackage{amssymb,amsmath,verbatim,mathtools,needspace,enumitem,etoolbox,graphicx,physics,microtype,afterpage,xspace,tabularx,lmodern,multirow,bm}
\usepackage{gensymb}
\usepackage[dvipsnames, usenames]{xcolor}
\definecolor{linkcolor}{rgb}{0.0,0.3,0.5}
\usepackage[unicode, colorlinks=true, linkcolor=linkcolor, citecolor=linkcolor, filecolor=linkcolor, urlcolor=linkcolor, linktocpage, breaklinks]{hyperref}
\usepackage[all]{hypcap}
\usepackage[T1]{fontenc}
\usepackage[utf8]{inputenc}
\usepackage[usenames,dvipsnames]{xcolor}
\usepackage{csquotes}
\usepackage{booktabs}
\hypersetup{colorlinks=true,citecolor=romared,linkcolor=romared,urlcolor=romared}

\definecolor{romared}{RGB}{142,0,28}

\newcommand{\be}{\begin{equation}}
\newcommand{\ee}{\end{equation}}

\def\be{\begin{equation}}
\def\ee{\end{equation}}
\newcommand{\beq}{\begin{eqnarray}}
\newcommand{\eeq}{\end{eqnarray}}
\usepackage{makecell}
\usepackage{soul}

\newcommand{\aap}{A\&A}
\newcommand{\mnras}{MNRAS}

\newcolumntype{Y}{>{\centering\arraybackslash}X}

\usepackage{ulem}

\begin{document}


\title{Constraining accretion physics with gravitational waves from \\  eccentric extreme-mass-ratio inspirals}

\author{Francisco Duque} 
\affiliation{Max Planck Institute for Gravitational Physics (Albert Einstein Institute) Am Mühlenberg 1, D-14476 Potsdam, Germany}
\author{Shubham Kejriwal} 
\affiliation{Department of Physics, National University of Singapore, Singapore 117551}
\author{Laura Sberna}
\affiliation{School of Mathematical Sciences, University of Nottingham, University Park, Nottingham NG7 2RD, United Kingdom}
\author{Lorenzo Speri}
\affiliation{European Space Agency (ESA), European Space Research and Technology Centre (ESTEC), Keplerlaan 1, 2201 AZ Noordwijk, The Netherlands}
\affiliation{Max Planck Institute for Gravitational Physics (Albert Einstein Institute) Am Mühlenberg 1, D-14476 Potsdam, Germany}
\author{Jonathan Gair}
\affiliation{Max Planck Institute for Gravitational Physics (Albert Einstein Institute) Am Mühlenberg 1, D-14476 Potsdam, Germany}

\begin{abstract}
We study the evolution of eccentric, equatorial extreme-mass-ratio inspirals (EMRIs) immersed in the accretion disks of active galactic nuclei. We find that single gravitational-wave observations from these systems could provide measurements with $\sim 10 \%$ relative precision of, simultaneously, the disk viscosity and mass accretion rate of the central supermassive black hole. 
This is possible when the EMRI transitions, within the observation time, from supersonic to subsonic motion relative to the disk gas, for eccentricities $e \gtrsim 0.025-0.1$. The estimate of the accretion rate would assist in the identification of the EMRI's host galaxy, or the observation of a direct electromagnetic counterpart, improving the chances of using these sources as cosmological sirens. Our work highlights the rich phenomenology of binary evolution in astrophysical environments and the need to improve the modelling and analysis of these systems for future gravitational-wave astronomy.
\end{abstract}

\maketitle 

\section{Introduction}\label{sec:intro}
The role that astrophysical environments play in binary coalescences has gained increasing attention in recent years. Accretion disks~\cite{Kocsis:2011dr, Barausse:2014tra, Derdzinski:2020wlw, Zwick:2021dlg, Zwick:2022dih, Speri:2022upm, 2020ApJ...892...90C, Garg:2022nko, Barausse:2007dy, Duffell:2024fwy, Zwick:2024yzh, Lyu:2024gnk}, dark matter distributions~\cite{Speeney:2022ryg, Cardoso:2021wlq, Cardoso:2022whc, Brito:2023pyl, Cannizzaro:2023jle, Destounis:2022obl, Baumann:2021fkf, Baumann:2022pkl, Tomaselli:2023ysb, Bamber:2022pbs, Aurrekoetxea:2023jwk, Tomaselli:2024bdd, Duque:2023cac, Tomaselli:2024bdd, Boskovic:2024fga, Rahman:2023sof, Tomaselli:2024ojz} and third-body companions~\cite{Cardoso:2021vjq, Gupta:2019unn, Suzuki:2020zbg, Sberna:2022qbn, Gupta:2022fbe, Oancea:2023hgu, Maeda:2023tao, Zwick:2024yzh, Yin:2024nyz, Camilloni:2023rra} all lead to distinctive features in the trajectory of a binary, and, consequently, in its gravitational wave (GW) signature~\cite{Cole:2022yzw, Zwick:2022dih, Kejriwal:2023djc, Santoro:2024tmo, Ghosh:2024arw}. 

Extreme-mass-ratio inspirals (EMRIs) are particularly sensitive to environmental effects. EMRIs are composed by a \textit{secondary} stellar-mass compact object, typically a black hole (BH) with mass $\sim 10-100\, M_\odot$, orbiting around a \textit{primary} supermassive black hole (SMBH) with mass $\gtrsim 10^5\, M_\odot$.
They are one of the main scientific targets of the Laser Interferometer Space Antenna (LISA)~\cite{2017PhRvD..95j3012B, LISA:2022yao, LISA:2022kgy, LISAConsortiumWaveformWorkingGroup:2023arg, Colpi:2024xhw}, a space-based GW interferometer sensitive to the mHz frequency band, which has recently been adopted by the European Space Agency and is currently planned to be launched in December 2035. EMRIs will complete thousands of orbital cycles while in the LISA observation window. Even small deviations from vacuum general relativity (GR), like the ones introduced by the presence of an environment, can become distinguishable when accumulated over so many orbits~\cite{Speri:2022upm, Cole:2022yzw, Barsanti:2022vvl, Kejriwal:2023djc, Speri:2024qak, Lyu:2024gnk}.

Most studies of environmental effects for LISA sources have been restricted to circular orbits. However, EMRIs formed via multibody scattering in dense nuclear clusters are expected to enter the observation band with a large eccentricity and retain it up to coalescence~\cite{2017PhRvD..95j3012B, Bonetti:2020jku, Vazquez-Aceves:2021xwl}. The interaction with gas-rich environments, like the accretion disk of an active galactic nucleus (AGN), also assists EMRI formation and increases their rates per SMBH by orders of magnitude~\cite{Pan:2021oob, Pan:2021ksp, Wang:2022obu, Derdzinski:2022ltb}. In fact, the collision between the secondary and the disk may be the mechanism behind the 
Quasi-Periodic Eruptions/Oscillations, emitting in the soft X-ray band (\textcolor{black}{with typical disk temperatures $\sim 10^6 \, \text{K}$ or} $\sim 0.1 \, \text{keV}$)~\cite{Kejriwal:2024bna, 2023ApJ...957...34L, 2023A&A...675A.100F, Arcodia:2024taw, Zajacek:2024kcw}. EMRIs formed via this \textit{wet} channel are expected to be equatorial and have negligible eccentricities when entering the LISA band~\cite{Tanaka_2002, Tanaka_2004, Pan:2021oob, Pan:2021ksp, Zwick:2021ayf}. However, eccentricity damping might be reduced when considering the entire evolution history of these EMRIs, e.g., when considering interactions with other compact bodies or disk turbulence, or the fact that accretion by the SMBH can have quiescent periods where the disk density is smaller~\cite{Cardoso:2020iji, Hu:2023oiu, Zwick:2021dlg, Gupta:2019unn, Suzuki:2020zbg, Gupta:2022fbe, Oancea:2023hgu, Maeda:2023tao, 2005A&A...443.1067N, Johnson:2006eh, 2009ApJ...707.1233Y, 2022MNRAS.513.5465S, 2019A&A...621A.101B, Franchini:2024pgl, Tomaselli:2024bdd, 2024MNRAS.528.4958W, 10.1093/mnrasl/slad183, Ogihara:2006dn, 2024MNRAS.528.4958W}. 
LISA is expected to measure the eccentricity of EMRIs at coalescence with absolute precision of $~10^{-5}$~\cite{2017PhRvD..95j3012B, Colpi:2024xhw}, so it is timely to extend studies of these environmental effects to eccentric inspirals. 

An EMRI evolving in an AGN disk is subject to several effects~\cite{binney2011galactic}: mass \textit{accretion} by both the primary and the secondary~\cite{1944MNRAS.104..273B, 10.1093/mnras/112.2.195, Kocsis:2011dr, Barausse:2014tra}; axisymmetric gravitational effects from the disk's self gravity~\cite{Kocsis:2011dr, Barausse:2014tra}; hydrodynamical drag induced by the relative velocity between the secondary and the gas in the disk~\cite{Barausse:2007dy, Kocsis:2011dr, Barausse:2014tra}; \textit{dynamical friction} by a density wake generated by the secondary's gravity, which then trails behind the secondary and exerts an additional gravitational pull on it~\cite{1943ApJ....97..255C, Ostriker:1998fa, Vicente:2019ilr}. For a circular binary, this wake forms a spiral density wave assisted by the disk's differential rotation, which can resonantly exchange angular momentum with the secondary~\cite{1980ApJ...241..425G, Tanaka_2002, Tanaka_2004, Kocsis:2011dr, Hirata:2010vn, Hirata:2010vp, 2011ApJ...737...37M,  2020MNRAS.494.5666I, Gilbaum:2024xxq}. This effect is analogous to (planetary) \textit{migration} in protoplanetary disks. 

Past work showed that of all the effects listed above, migration is the most relevant for quasicircular EMRIs where the secondary is embedded in an AGN disk~\cite{Kocsis:2011dr, Barausse:2014tra, Speri:2022upm}. Using state-of-the-art waveform models for EMRIs~\cite{Chua:2020stf, Katz:2021yft} and a Bayesian framework for parameter estimation, Speri \textit{et al}.~\cite{Speri:2022upm} demonstrated that LISA would be able to measure migration torques in different accretion scenarios and place constraints on the disk properties (e.g. viscosity, accretion rate), if GW observations are combined with multimessenger electromagnetic counterparts. If migration is neglected in the analysis of these systems, one incurs biases in the mass and spin of the primary that may jeopardize tests of fundamental physics and lead to false deviations from GR~\cite{Kejriwal:2023djc}.

However, migration arises from resonances between the secondary's orbital frequency and epicyclic motion at different radii of the disk. This is a \textit{global} effect, where one needs to consider density/temperature gradients along the extended gas distribution. 
 As eccentricity increases, these resonances become less effective~\cite{Zwick:2021ayf, 2011ApJ...737...37M,  2020MNRAS.494.5666I}. In fact, binaries with eccentricities $e \gtrsim 0.025-0.1$ become supersonic relative to the (local) gas in the disk. The transition from subsonic to supersonic motion alters the nature of the secondary-disk interactions. In the subsonic regime, the front and back wake around the secondary are almost symmetric, which cancels their net effect, and it is necessary to consider the interaction with gas further away from the secondary~\cite{1980ApJ...240...20R, Ostriker:1998fa, Vicente:2019ilr}. As the secondary transitions to the supersonic regime, this symmetry is broken, the spiral structure typical of migration does not form, and \textit{local} dynamical friction becomes the dominant environmental effect~\cite{10.1046/j.1365-8711.2000.03466.x, 2008A&A...482..677C, 2010A&A...523A..30B, 2011ApJ...737...37M,  2020MNRAS.494.5666I, Sánchez-Salcedo_2019, 2020ApJ...897..142S}. 

In this work, we extend the study of EMRIs in accretion flows to eccentric orbits, equatorial and prograde with respect to the motion of the gas in the disk. In Section~\ref{sec:Setup}, we construct a phenomenological model describing the secular correction in the secondary's trajectory due to the interactions with the gas environment, including the eccentricity evolution. Our model is inspired by previous explorations in the context of planetary migration, where hydrodynamical simulations are used to match analytic estimates in both the subsonic and supersonic limit~\cite{2011ApJ...737...37M, 2020MNRAS.494.5666I, 2019ApJ...885..152S}. As we will see, in the supersonic regime, eccentricity damping is weakened, and its typical evolution timescale \textcolor{black}{can become comparable and even larger than the inspiralling timescale}. 

We implement our model in the state-of-the-art \texttt{FastEMRIWaveform} (\texttt{few}) package for EMRI waveform computation~\cite{Chua:2020stf, Katz:2021yft, Speri:2023jte} and use a Fisher-Matrix framework to study inference with LISA in Section~\ref{sec:Constraints}. We find that the presence of eccentricity typically helps to probe the astrophysical environment of the EMRI, even for values as small as $e_0 \sim 0.01$ at the start of the observation. Moreover, it allows for independent measurements of both disk viscosity and the primary mass accretion rate, without the need for an electromagnetic counterpart, something that was not possible for observations of circular orbits. We also comment, in Section~\ref{sec:Formation} and~\ref{sec:Multimessenger}, on the implications of our results for EMRI formation in gaseous environments, and multimessenger astronomy with EMRIs, respectively. Finally, in Section~\ref{sec:Conclusions} we discuss the limitations of our work and identify challenges for the future. Unless stated otherwise, we adopt geometric units with $G=c=1$.

\section{Methods}\label{sec:Setup}
In this Section, we summarize the adopted disk model and construct a phenomological model for the interaction of the EMRI with the disk environment, valid up to large eccentricities. We also describe how to implement these effects in \texttt{few} for waveform evaluation and review the Fisher-matrix framework used for inference.

\subsection{Accretion-disk model}

We denote the mass of the primary central BH by $M$ and the mass of the secondary compact body by $\mu$. Following Refs.~\cite{Zwick:2021dlg, Derdzinski:2020wlw}, we consider a Newtonian, radiatively-efficient, stationary, thin-accretion disk, that is accreted by the central SMBH at a steady-state rate
\beq
\dot{M}\sim 3\pi \nu \Sigma \sim \text{constant} \, ,
\eeq
with $\Sigma$ the disk's surface density and $\nu$ its kinematic viscosity. The viscosity is taken to be parametrized by a constant $\alpha$, such that $\nu = \alpha c_s H$, where $H$ is the disk scale height, $c_s = H \Omega_K$ the isothermal sound speed, and $\Omega_K = \sqrt{M/r^3}$ the Keplerian orbital frequency. The disk density is then $\rho = \Sigma / 2 H$ and the thin-disk approximation needs to satisfy $2H \ll r$ (this is verified throughout the evolution of the EMRI). 

We parameterize the disk profile according to a power law for the surface density $\Sigma(r)$ and the aspect ratio $h(r)$
\beq
\Sigma(r) &=& \Sigma_0 \left(\frac{r}{10M} \right)^{-\Sigma_p} \, , \label{eq:PowerLawSurface} \\
h(r) &=& \frac{H(r)}{r} = h_0 \left(\frac{r}{10M} \right)^{(2\Sigma_p-1)/4} \, , \label{eq:PowerLawAspectRatio} 
\eeq
where $\Sigma_p \in \mathbb{Z} $, $\Sigma_0$ and $h_0$ are typical values in the inner region of the disk
, and the power appearing in $h(r)$ ensures the primary accretes mass at a constant rate. We adopt this parameterization since it allows to quickly identify when the secondary is in the subsonic or supersonic regime by comparing the orbit's eccentricity $e$ with $h_0$.  

Our disk model reformulates 
the Shakura-Sunayev $\alpha$-model~\cite{Shakura:1972te, Abramowicz:2011xu, frank_king_raine_2002, Derdzinski:2022ltb}, which captures the general features observed in magnetohydrodynamic simulations of turbulent, radiatively-efficient disks~\cite{Nouri:2023nss, Chen:2023shf, Gangardt:2024bic}. The $\alpha$-disk model takes the viscous stress to be proportional to the total pressure in the disk, i.e. $t_{r\varphi}= \alpha \left(p_\text{gas} + p_\text{rad} \right)$, where $p_\text{gas}$ is the thermal gas pressure and $p_\text{rad}$ is the radiation pressure. Observations and numerical simulations indicate that in the disk's inner region ($r \lesssim 10^2 M$), pressure is dominated by radiation and opacity by electron scattering~\cite{1994ApJ...427..987B, 2003MNRAS.341..501S, Derdzinski:2022ltb}. Under these assumptions~\cite{Shakura:1972te, Kocsis:2011dr, Speri:2022upm, Garg:2022nko}
\beq
\Sigma_\alpha &\approx& 5.4 \times 10^2 \left( \frac{0.1}{\alpha} \right) \left(\frac{0.1}{f_\text{Edd}}\frac{\epsilon}{0.1} \right)\Big(\frac{r}{10M} \Big)^{3/2} \left[ \frac{\text{g}}{\text{cm}^2} \right]  \, , \nonumber \\ \label{eq:Shakura-Sunayev_Sigma} \\
h_\alpha &\approx& 0.15\left(\frac{f_\text{Edd}}{0.1}\frac{0.1}{\epsilon} \right)\Big(\frac{r}{10M}\Big)^{-1} \, .
\label{eq:Shakura-Sunayev_h}
\eeq
Here, $f_\text{Edd}$ is the fraction of the Eddington rate at which the primary is accreting and $\epsilon$ is the efficiency of conversion of mass-energy to radiation, such that $\dot{M}= f_\text{Edd} L_\text{Edd}/\epsilon $, with $L_\text{Edd}$ the Eddington luminosity. These quantities appear in the disk structure always through the ratio $f_\text{Edd} / \epsilon$. Their typical values are $f_\text{Edd} \sim 0.01-0.1$, $\epsilon \sim 0.1$, and for the viscosity coefficient $\alpha \sim 0.001 - 0.1$~\cite{Jiang:2019bxn, 2010ApJ...713...52D, Nouri:2023nss}. These values correspond to $\Sigma \sim 10^3 - 10^5 \, \text{g}/\text{cm}^2$ and $h \sim 0.01 - 0.1$, which are compatible with current observations~\cite{Jiang:2019ztr}. 

By picking $\Sigma_p = -3/2$ in the power-law model~\eqref{eq:PowerLawSurface}-\eqref{eq:PowerLawAspectRatio}, any choice of $\Sigma_0$ and $h_0$ can then be mapped to a combination of $\alpha$ and $f_\text{Edd} / \epsilon$ of the $\alpha$-model. We will focus on this profile in this work, but one can repeat the mapping for any particular disk solution with different physics, like a different opacity law governing a different region of the disk (see Table~1 and Fig.~1 of Ref.~\cite{Derdzinski:2022ltb}) or a different dependence between the viscous stress and pressure. \textcolor{black}{For example, for the outer region of the $\alpha$-disk ($r \gtrsim 10^2M$), pressure becomes dominated by gas and the profile is well approximated by $\Sigma_\alpha \sim r^{-1/2}$, while $h \sim \text{constant}$ (up to $r \lesssim 10^4M$). Another example} is the $\beta$-disk model, where viscous stress is proportional to only the thermal gas pressure~\cite{Shakura:1972te, Abramowicz:2011xu, frank_king_raine_2002}, i.e. $t_{r\varphi}= \alpha p_\text{gas}$. This choice cures the thermal instabilities present in the $\alpha$-disk and leads to $\Sigma_\beta \propto r^{-3/5}$, but is less favoured by both simulations and observations. Ultimately, we can remain agnostic about the internal physical properties of the disk and ask instead what regions of the parameter space $\left\{\Sigma_0, \,  h_0,\, \Sigma_p \right\}$ lead to distinguishable environmental effects in the GW signature of an EMRI. Though some of them may not directly correspond to known disk solutions, they might still be physically relevant considering all the uncertainties in the astrophysics of AGN disks, the discrepancies between the finer details of numerical simulations and simple analytical models like the $\alpha$-disk, and the time-dependence in the disk structure due to the inspiral of the secondary and instabilities (e.g. turbulence)~\cite{Jiang:2019bxn, Nouri:2023nss, Derdzinski:2020wlw, Gangardt:2024bic}. 

 We are also ignoring relativistic corrections~\cite{Novikov:1973kta} and changes at the inner edge of the disk, which would complicate the power-law profiles proposed in Eqs.~\eqref{eq:PowerLawSurface}-\eqref{eq:PowerLawAspectRatio}. Imposing that viscous stress vanish at the disk's inner edge at some radius $r_\text{in}$ introduces corrective factors of
\beq
F^4= 1-\sqrt{r_\text{in}/r} \, ,\label{eq:FudgeFactor}
\eeq
up to some power in the disk structure (e.g. $\Sigma_\alpha \propto F^{-1}$ and $h_{\alpha} \propto F^4$)~\cite{Shakura:1972te, Kocsis:2011dr, Speri:2022upm, Garg:2022nko}. Since we are mostly interested in how eccentricity affects the interaction of the secondary with the disk, we ignore these corrections. 

\subsection{Torque  model }\label{sec:TorqueModel}

The relative velocity $\Delta v$ between the secondary and the (local) gas in the disk dictates the nature of their gravitational interaction. For equatorial orbits prograde with respect to the disk rotation
\beq
\Delta v \approx e  v_K = e r \Omega_K 
= \frac{e}{h}c_s \, , 
\eeq
where $v_K$ is the Keplerian velocity at radius $r$. Therefore, for $e \lesssim h$, the EMRI is in the subsonic regime and vice-versa~\cite{2011ApJ...737...37M, 2020MNRAS.494.5666I}. Considering that $h \gtrsim 0.02$ in the inner disk region, an EMRI can attain supersonic motion even with moderately low eccentricity.  

\subsubsection{Subsonic motion}

In the quasicircular, subsonic regime, the EMRI excites spiral density waves due to the shear motion of the disk gas~\cite{1980ApJ...241..425G, Tanaka_2002, Tanaka_2004, Kocsis:2011dr, Hirata:2010vn, Hirata:2010vp, 2011ApJ...737...37M,  2020MNRAS.494.5666I}. 
The wake then backreacts on the binary by exerting additional forces/torques, which excite resonances at specific locations in the disk (e.g., Lindblad resonances). As a consequence, the binary ``migrates'', either inwards or outwards depending on the net torque between ingoing spiral waves, which transfer angular momentum to the binary, and outgoing ones, which extract angular momentum from it. Migration can be understood as dynamical friction, but takes into account the differential rotation of the disk gas at radii far-away from the secondary. If the secondary is massive enough, it can open a gap in the disk and migration is said to be of Type-II. 
The EMRIs we will study \textit{do not} fulfill this condition~\cite{Kocsis:2008aa, Gilbaum:2024xxq}, and thus we consider only Type-I migration. 

In general, migration torques act on the inverse timescale 
\beq
\frac{1}{t_\text{gas}} = \varepsilon \left(\frac{\Sigma r^2}{M} \right)\frac{\Omega_K}{h^4} \, , 
\eeq
%
where $\varepsilon = \mu / M$ is the small mass ratio~\footnote{to not be confused with the mass-radiation conversion efficency $\epsilon$.}. Ref.~\cite{Tanaka_2002, Tanaka_2004} used linear perturbation theory to compute the force felt by the secondary at every location in a slightly eccentric prograde orbit with $e \ll h$, taking into account corotation torques and the 3D structure of the disk (they also included inclination in their analysis, but we focus on equatorial orbits). This force can then be averaged over one period to obtain the timescales for the evolution of the orbital parameters, namely the semi-major axis $a_\text{m}$ (or alternatively the semi-latus rectum $p$) and eccentricity
\beq
\frac{1}{t^\text{sub}_{a_\text{m}}} &=& -\frac{\left< da_\text{m}/dt \right>}{a} = 2\frac{C_\text{sub}}{t_\text{gas}} h^2 \, , \label{eq:asubsonic}\\ 
\frac{1}{t^\text{sub}_e} &=& -\frac{\left< de/dt \right>}{e}  = \frac{0.78}{t_\text{gas}} \, ,  \label{eq:esubsonic}
\eeq
where the radial-dependent quantities are to be evaluated at $a_\text{m}$. $C_\text{sub}$ is a numerical factor that may depend on the density and temperature gradient of the disk.
%
%
It is calibrated with hydrodynamical simulations~\cite{2017MNRAS.471.4917J}, but these estimates still have uncertainties and do not take into account all features present in AGN disks, such as:
\begin{enumerate}
    \item Reflections of the wake at the disk's inner edge, which are expected to weaken the torque when the reflected wave reaches the secondary~\cite{2011ApJ...741..109T, 2018MNRAS.473.5267M};
    \item Increase in the surface density, which increases the torques, at the disk's inner edge introduced by the $F$ factor~\eqref{eq:FudgeFactor}; 
    \item Magnetic fields, which can generate turbulent flows~\cite{10.1093/mnras/stw843, Terquem:2003vt, Fromang:2005dk, Uribe:2011ag, 2008ApJ...679..813M, Sanchez-Salcedo:2011wwf, Dhang:2024aoa, Wu:2023qeh}, responsible for stochastic torque fluctuations. These can also arise from asymmetries in the gas flow near the secondary~\cite{2005A&A...443.1067N, Johnson:2006eh, Kley:2012ue,   Derdzinski:2020wlw, Zwick:2021dlg}. Stochastic torques lead to ``diffusive'' migration similar to a random walk, and typically increase eccentricity and inclination~\cite{2005A&A...443.1067N, Ogihara:2006dn}. If the turbulence is too strong, it can even lead to the disruption of the spiral flow patterns, severely hindering Lindblad resonances, and consequently type-I migration based on linear estimates~\cite{Wu:2023qeh};
    \item Radiative effects from relaxing the disk isothermality condition~\cite{2017MNRAS.471.4917J}, which in turn relaxes the relation between the power-law for the aspect ratio~\eqref{eq:PowerLawAspectRatio} and the one for the surface density~\eqref{eq:PowerLawSurface}. 
    \item ``Thermal'' torques from heat diffusion near the secondary~\cite{10.1093/mnras/stx2271, Hankla_2020, Grishin:2023riv, Gangardt:2024bic}. In many disk regions, thermal torques have a positive sign and thus lead to outward migration. They can dominate over gravitational torques (and radiation reaction from GWs) in outer regions of the disk ($r \gtrsim 800 M$), and even create \textit{migration traps}, where the net torque felt by the secondary is zero. Thermal torques also typically increase eccentricity and inclination~\cite{10.1093/mnras/stz718}. 
    \item Relativistic effects both in the disk structure and wake dynamics~\cite{Hirata:2010vn, Hirata:2010vp}. 
\end{enumerate}

The corrections listed above mostly change the value of $C_\text{sub}$ by a numerical factor of $\mathcal{O}(1)$. Though this variation can impact parameter estimation, in this work we are more interested in understanding the role that eccentricity plays in the detectability of environmental effects. 
\textcolor{black}{We use the most recent results in the literature \cite{2024ApJ...968...28T}}\footnote{\textcolor{black}{These estimates include the radial gradient of the disk temperature,  which for steady-state accretion satisfies $T_p = - d \log T/ d r = 3/2 - \Sigma_p$.  } }
\beq
C_\text{sub} = 2.15 + 0.04 \Sigma_p  \, .
\eeq
Note that these modelling uncertainties should also correspond to different numerical factors for the amplitude of $1/t_e$, but these have been much less explored. We therefore use the most common value found in the literature ($0.78$)~\cite{Tanaka_2004, 2020MNRAS.494.5666I}.

Finally, as mentioned above, we focus on prograde motion of the secondary with respect to the gas flow in the disk. The evolution timescales for retrograde motion can differ by orders of magnitude~\cite{Sánchez-Salcedo_2018, Sánchez-Salcedo_2020} -- and also lead to eccentricity excitation -- but according to formation models, EMRIs immersed in AGN disks will be already prograde when they enter the LISA band~\cite{2024MNRAS.528.4958W, Stone:2016wzz, Derdzinski:2022ltb}. 

\subsubsection{Supersonic motion}
%
\begin{figure*}[t!]
\centering
\includegraphics[width=0.49\linewidth]{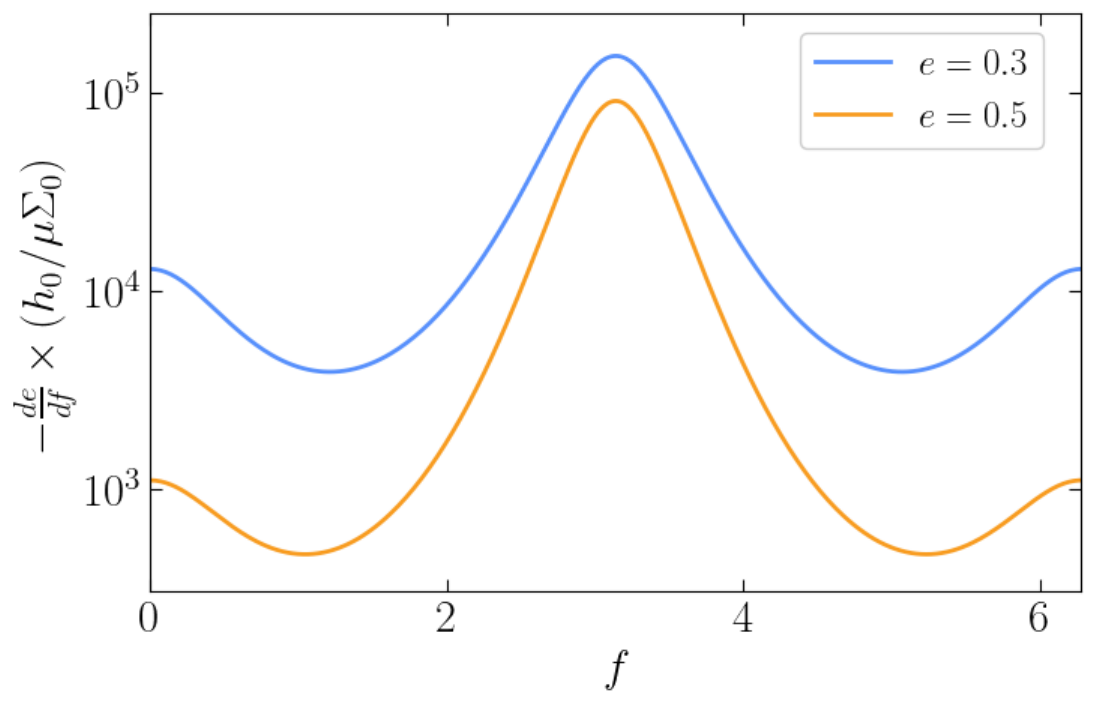}\hspace{5pt}
\includegraphics[width=0.49\linewidth]{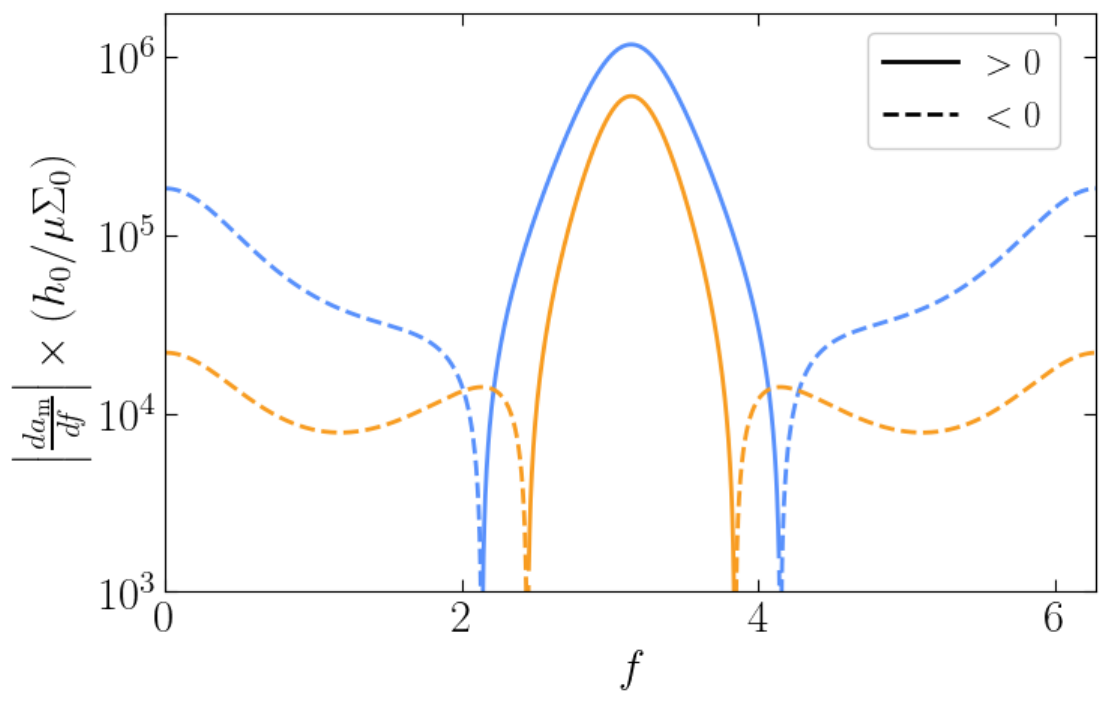} 
\caption{\textit{Left:} Variation of the eccentricity over one complete orbit due to dynamical friction in the inner region of an $\alpha$-disk ($\Sigma_p = -3/2$ for the power-law surface density in Eq.~\eqref{eq:PowerLawSurface}; typically, $\Sigma_0/h_0 \sim 10^{-15}-10^{-13}$ in geometrized units). The secondary is in supersonic motion relative to the gas flow on an orbit with semi-major axis $a_\text{m}=10M$. Dynamical friction always acts to circularize the orbit ($de/df < 0$ during the whole orbit); \textit{Right:} same as the left panel but for the semi-major axis. The dashed segments correspond to negative $da_\text{m}/df$. At the periapsis (when the true anomaly $f=0, \, 2\pi$), dynamical friction is negative and the secondary feels a headwind that takes energy away from the binary. At the apoapsis ($f=
\pi$), the opposite occurs and the interaction with the environment is acting as a thrust. In this disk region, the surface density increases with radius, so the contribution at apoapsis will dominate and the net dynamical friction is outwards, decelerating the inspiral.}
\label{fig:VariationOneOrbit}
\end{figure*}

For larger orbital eccentricities, the EMRI motion relative to the disk becomes supersonic and the gas shear flow becomes less effective. In this case, dynamical friction from the local gas around the secondary becomes the dominant environmental effect. To compute it we employ the so-called local approximation~\cite{Sánchez-Salcedo_2019}, which assumes that:
\begin{enumerate}
    \item Dominant contributions to the drag force exerted by the disk come from regions close enough to the secondary, such that density and temperature gradients can be ignored to compute the density wake;
    \item Differential rotation of the disk is negligible;
    \item Curvature of the wake is negligible. In particular, the fact that the secondary can catch its own wake is ignored;
\end{enumerate}

In practice, the drag force is computed assuming (recti)linear motion and a constant density disk. Using the same strategy of Tanaka \textit{et} al.~\cite{Tanaka_2002, Tanaka_2004}, this force can then be averaged over one orbit to obtain the typical timescales for the evolution of the orbital parameters. This approach has been validated with hydrodynamical simulations for extreme-mass ratios~\cite{2019ApJ...885..152S, 2020MNRAS.494.5666I}. 

In the supersonic limit, the drag force felt by a small body in a slab-like geometry asymptotes to the time-independent value~\cite{2011ApJ...737...37M, 2013ApJ...762...21C, Vicente:2019ilr}
\beq
\mathcal{F}_\text{DF} &=& - 2\pi \mu^2 \frac{\Sigma}{2H} \frac{1}{\Delta v^2}\eta \, , \label{eq:DFSuper}
\eeq
where $\eta$ is a softening parameter that takes into account the effective size of the secondary, since in the supersonic regime the force diverges at the location of the point particle. This is typically written in terms of a Couloumb logarithm, with Ref.~\cite{2013ApJ...762...21C} proposing  
\beq
\eta = \log \left( \frac{7.15H}{r_\text{acc}} \right) \, , 
\eeq
where $r_\text{acc} = 2 \mu / \Delta v$ is the accretion radius of the secondary. In general, this quantity varies along the trajectory, which complicates the dependence of the drag force~\eqref{eq:DFSuper} on the orbital parameters. However, since the dependence is only logarithmic, the change in $\mathcal{F}_\text{DF}$ should be small. Therefore, to simplify the model, we take $\eta = 1$ during the inspiral. Note also the drag force is independent of the sound speed/viscosity and thus is independent of the thermal properties of the disk.



To then compute the averaged effect this drag force has on equatorial, eccentric, prograde inspirals we repeat the steps followed in Ref.~\cite{2008ApJ...679..813M} (for Keplerian orbits). 
We take cylindrical coordinates $\left( r, \, \varphi, \, z\right)$ with origin at the SMBH. The velocity of the secondary is
\beq
\bm{v}_p =\sqrt{\frac{M}{p}}e\sin f \bm{e}_r + \sqrt{\frac{M}{p} } \left(e \cos f + 1 \right) \bm{e}_\varphi \, , 
\eeq
while the gas velocity at distance $r$ from the primary is
\beq
\bm{v}_g = \sqrt{\frac{M}{r}}  \bm{e}_\varphi = \sqrt{\frac{M(1+e \cos f)}{p}}  \bm{e}_\varphi \, , 
\eeq
where $\bm{e}_r$ and $\bm{e}_\varphi$ are, respectively, the radial and azimuthal unit vectors, $p$ is the semi-latus rectum and $f$ is the true anomaly. The relative velocity between the secondary and the gas in the accretion disk is thus
\beq
\Delta  \bm{v} &=& \bm{v}_p - \bm{v}_g = \Delta v_r \bm{e}_r +  \Delta v_\varphi \bm{e}_\varphi \, , \\
\Delta v_r &=& \sqrt{\frac{M}{p}}e\sin f  \, , \\
\Delta v_\varphi &=&  \sqrt{\frac{M}{p} } \left(e \cos f + 1 \right)  - \sqrt{\frac{M(1-e^2)}{p}}  \, , 
\eeq
and it determines the direction of the drag force 
\beq
\bm{\mathcal{F}}_\text{DF} = \mathcal{F}_\text{DF} \left( \frac{\Delta v_r}{\Delta v} \bm{e}_r + \frac{\Delta v_\varphi}{\Delta v} \bm{e}_\varphi \right)\, .
\eeq

Now consider a generic perturbing force $\bm{\mathcal{F}} = \mathcal{R} \bm{e}_r + \mathcal{S} \bm{e}_\varphi $ to the Kepler problem. The evolution of the orbital parameters is governed by the osculating equations (also called Gauss or Lagrange planetary equations)~\cite{poisson_will_2014, Gair:2010iv} 
\beq
\frac{dp}{dt} &=& 2\sqrt{\frac{p^3}{M}}\frac{1}{1+e\cos f}\, \mathcal{S} \, , \\
\frac{de}{dt} &=&\sqrt{\frac{p}{M}}\left[\sin f \, \mathcal{R} + \frac{2\cos f+ e(1+\cos^2 f)}{1+e \cos f}\,\mathcal{S} \, \right] , \\
\frac{df}{dt} &=& \sqrt{\frac{M}{p^3}}\left( 1 + e \cos f \right)^2 \nonumber \\ &&+ \frac{1}{e}\sqrt{\frac{p}{M}}\left[ \cos f \, \mathcal{R} - \frac{2 + e \cos f}{1 + e \cos f}  \sin f \, \mathcal{S} \right]\, .  
\eeq
Assuming the perturbing force is small compared to the gravitational force, one can take the orbital elements to be varying slowly and fix their values on the right-hand side to be constant during one orbit. Rewriting the osculating equations in terms of $f$ at leading order in the perturbing force 
\beq
\frac{dp}{df} &\approx & 2\frac{p^3}{M}\frac{1}{(1+e\cos f)^ 3}\, \mathcal{S} \, ,\label{eq:dxdf1} \\
\frac{de}{df} &\approx & \frac{p^2}{M}\left[\frac{\sin f}{(1+e\cos f)^2} \, \mathcal{R} + \frac{2\cos f+ e(1+\cos^2 f)}{(1+e \cos f) ^3}\,\mathcal{S} \, \right] ,\nonumber \\ \label{eq:dxdf2}\\
\frac{dt}{df} &\approx & \sqrt{ \frac{p^3}{M} }\frac{1}{(1+e \cos f)^2}\times \Bigg[1 - \frac{p^2}{e \, M} \times \nonumber \\
&\times& \Bigg( \frac{ \cos f}{(1 + e \cos f)^2} \mathcal{R} - \frac{2 + e \cos f}{(1 + e \cos f)^3}  \sin f \, \mathcal{S} \, \Bigg) \Bigg] \, . \label{eq:dxdf3}
\eeq
The orbit-averaged quantities can be simply obtained by, e.g.
\beq
\left< \frac{dp}{dt} \right> = \frac{1}{T}\int_0^T dt \left(\frac{dp}{dt} \right) = \frac{1}{T}\int_0^{2\pi} df \left( \frac{dp}{df} \,  \right) \, ,
\eeq
where $T$ is the orbital period, and an analogous expression holds for eccentricity. The timescale of evolution of $p$ and the eccentricity $e$ can then be related with the one of $a_\text{m}$ through
\beq
a_\text{m} = \frac{p}{1-e^2} \, , 
\eeq
which implies
\beq
\left< \frac{da_\text{m}}{dt} \right> = \frac{1}{1-e^2} \left< \frac{dp}{dt} \right> + \frac{2\, p \,e}{(1-e^2)^2} \left< \frac{de}{dt} \right> \, . \label{eq:arelpe}
\eeq

In Fig.~\ref{fig:VariationOneOrbit}, we show the variation of $da_\text{m}/df$ and $de/df$ along one orbit for the disk model with $\Sigma_p = -3/2$, appropriate to describe the inner region of $\alpha$-disks ($r \lesssim 100M$). Our results are qualitatively 
similar to previous works~\cite{2011ApJ...737...37M}, though these studies never use the same disk profile as ours. Note that $da_\text{m}/df$ changes sign throughout the orbit. At the periapsis (when the true anomaly $f=0, \, 2\pi$), its value is negative, indicating that dynamical friction is extracting energy from the binary. The secondary feels a ``headwind'' and is being decelerated by the gas. At the apoapsis ($f=\pi$), the opposite occurs, and the gas flow is thrusting the secondary, accelerating it. Since the surface density of our disk model increases with radius, the contribution at the apoapsis dominates, and the net effect of dynamical friction over one orbit will result in outward migration (i.e. positive $\left< da_\text{m}/dt \right>$). \textcolor{black}{Note, however, that in this disk region, the orbital evolution is already dominated by GW emission, which still leads to an inspiral (i.e. negative $\left< da_\text{m}/dt \right>$ when both the GW and dynamical friction terms are considered). For outer regions of the disk ($r \gtrsim 100M$), the density decays with the radius and therefore the net dynamical friction is negative, leading to inward migration. We explore this outer disk region further in Appendix~\ref{app:OuterDisk}}. 

%
%
We obtain the following dependences for the orbit-averaged evolution of the orbital parameters 
\beq
&&\frac{1}{t^\text{sup}_{a_\text{m}}} \propto - \frac{\Sigma_0}{h_0}\frac{(1-e^2)}{e}a_\text{m}^3  \propto  \frac{1}{t_\text{gas}} \frac{h^3}{e}\left(1 - e^2\right)  \, , \label{eq:ta_sup} \\
&&\frac{1}{t^\text{sup}_e} \propto \frac
{\Sigma_0}{h_0}\frac{a_\text{m}^3}{e^3} \propto \frac{1}{t_\text{gas}}\left(\frac{h}{e}\right)^3 \, . \label{eq:te_sup}
\eeq
The proportionality factors can be found in Table~\ref{tab:Sub_vs_Super_Timescales}, \textcolor{black}{where we also list the factors for the outer disk region}. 

As the EMRI transitions from the supersonic to subsonic motion (or vice-versa), the functional dependence of the evolution timescales for $a_\text{m}$ and $e$ change. In the next section, we propose a simple phenomenological model for $1/t_{a_\text{m}}$ and $1/t_e$  that recovers the correct (analytic) behavior in the subsonic ($e \ll h$) and supersonic limit ($e \gg h$).

\subsubsection{Matching}
%
\begin{table*}[t]
    \small 
    \renewcommand{\arraystretch}{1.2} 
    \label{tab:Sub_vs_Super_Timescales}
    \begin{ruledtabular}
        \begin{tabular}{  @{\,} >{\centering\arraybackslash}m{2cm}  @{\,} | @{\,} >{\centering\arraybackslash}m{2.5cm} @{\,} |@{\,}  >{\centering\arraybackslash}m{3.5cm} @{\,} | >{\centering\arraybackslash}m{2.5cm} @{\,} | @{\,} >{\centering\arraybackslash}m{3cm} @{\,} | @{\,}  >{ \centering\arraybackslash}m{3cm}  @{\,}} 
            & Regime & Physical Effect & Disk Region & $t_\text{gas}/t_{a_\text{m}}$ & $t_\text{gas}/t_e$ \\
            \midrule
            \multirow{2}{=}{\centering $e \ll h$} 
            & \multirow{2}{=}{\centering Subsonic } 
            & \multirow{2}{=}{\centering Migration Torques} 
            & Inner & $4.2 h^2$ & \multirow{2}{=}{ \centering $0.78$} \\
            & & & Outer  & $4.3 h^2$ & \\
            \cmidrule{1-6}
            \multirow{2}{=}{ \centering $e \gg h$} 
            & \multirow{2}{=}{\centering Supersonic} 
            & \multirow{2}{=}{\centering Dynamical Friction} 
            & Inner  & $-28 (h^3/e)(1-e^2)$ & $23(h/e)^3$ \\
            & & & Outer  & $13 (h^3/e)(1-e^2)^{1/4}$ & $23(h/e)^3 (1-e^2)$  \\
        \end{tabular}
    \end{ruledtabular}
    \caption{
        Analytic estimates for the evolution timescales, \textcolor{black}{due to the interaction with the disk}, of the semi-major axis, $1/t_{a_\text{m}} = -\left<da_\text{m}/dt\right>/a_\text{m}$, and eccentricity, $1/t_e = -\left<de/dt\right>/e$, when the EMRI is in subsonic or supersonic motion relative to the gas in the accretion disk. \textcolor{black}{We show results for both the inner disk region ($r\lesssim 100M$), where $\Sigma_p = -3/2$ in Eqs.~\eqref{eq:PowerLawSurface}-\eqref{eq:PowerLawAspectRatio}, and the outer disk ($r\gtrsim 100M$), with  $\Sigma_p = 1/2$.}
    }
\end{table*}
\begin{figure*}[t]
\centering
\includegraphics[width=0.49\linewidth]{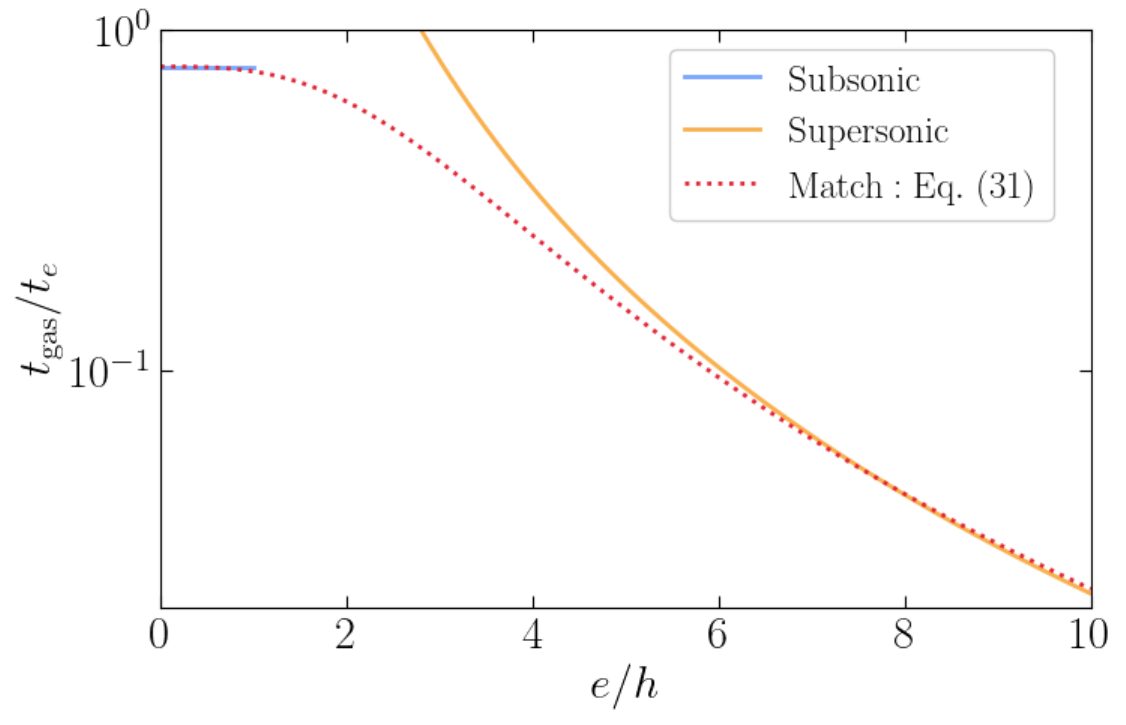}    
\hspace{1pt}
\includegraphics[width=0.49\linewidth]{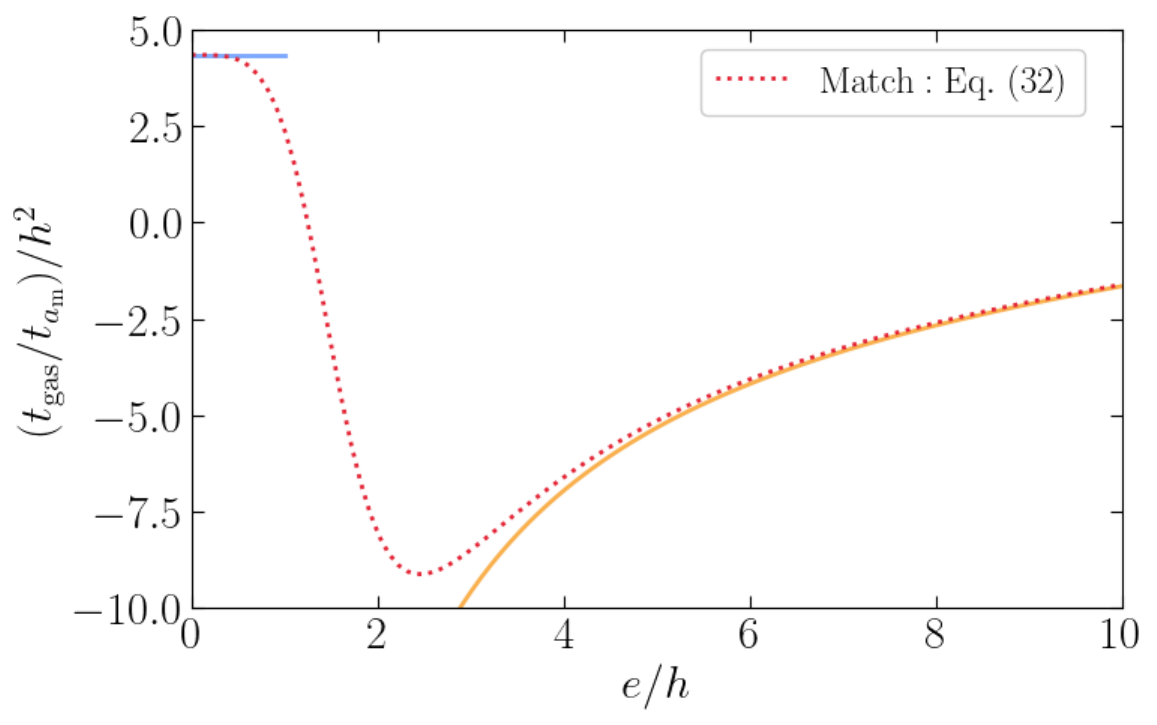}    
\caption{
Phenomenological model proposed in Eqs.~\eqref{eq:te_phenom}-\eqref{eq:ta_phenom} for the evolution timescales of the semi-major axis $a_\text{m}$ and eccentricity $e$ due to the interaction with the disk \textcolor{black}{in the inner disk region ($r \lesssim 100M$)}, which recovers the analytical estimates presented in Table~\ref{tab:Sub_vs_Super_Timescales} in their respective limits. This model is based on numerical fits to hydrodynamical simulations~\cite{10.1046/j.1365-8711.2000.03466.x, 2008A&A...482..677C, 2020MNRAS.494.5666I}.}
\label{fig:te_ta}
\end{figure*}

The results from the previous two subsections are summarized in Table~\ref{tab:Sub_vs_Super_Timescales}. When $e \sim h$, the linear analysis from dynamical friction is known to diverge due to the formation of shocks, and non-linear corrections become necessary~\cite{2011ApJ...737...37M, Vicente:2019ilr, Suzuguchi:2024btk, ONeill:2024tnl}.
To match the two regimes in this transonic region, we use a phenomenological model based on fits to hydrodynamical simulations~\cite{10.1046/j.1365-8711.2000.03466.x, 2008A&A...482..677C, 2020MNRAS.494.5666I}
\beq
\frac{t_\text{gas}}{t_e} &=& 0.78(1-e^2)^{1/4}x\Big/\left(1+\frac{1}{30}\left(\frac{e}{h}\right)^3 \right) ,  \label{eq:te_phenom} \\
\frac{t_\text{gas}}{t_{a_\text{m}}} &=& 2 C_\text{Sub}\, h^2 \left(1-e^2\right)\frac{1-\left(\frac{e}{1.25h}\right)^4}{1+\left(\frac{e}{1.75h}\right)^5} \, . \label{eq:ta_phenom}
\eeq
This model is illustrated in Fig.~\ref{fig:te_ta}, compared to the analytical estimates given in Table~\ref{tab:Sub_vs_Super_Timescales} for the subsonic and supersonic limits. \textcolor{black}{In Appendix~\ref{app:OuterDisk}, we give similar expressions and comparisons for the outer disk region ($r \gtrsim 100M$), where $\Sigma_p = 1/2$ in Eqs.~\eqref{eq:PowerLawSurface}-\eqref{eq:PowerLawAspectRatio}.}

One might question how the results presented in the next sections depend on the matching condition imposed for $e \sim h$, in particular, on the steepness of the transition or the presence of the bump in $1/t_{a_\text{m}}$ at $e \sim 2h$. We explore this in Appendix~\ref{app:Matching} and conclude that our results for the measurement precision of the disk parameters only differ by $\mathcal{O}(1)$, and the overall qualitative picture is left unchanged. We note, however, that the bump in the transition region of $1/t_{a_\text{m}}$ has been reported in several studies and is linked to the resonant excitation of sound waves when the relative velocity of the secondary with respect to the gas flow matches the sound speed, which occurs at $e\sim h$ (in particular see Fig.~1 in~\cite{2020MNRAS.494.5666I}). 

Also, recall that we are ignoring relativistic corrections while EMRIs inherently probe strong-field gravity. From previous studies~\cite{Barausse:2007ph, Traykova:2023qyv}, we expect relativistic corrections in dynamical friction to add a factor of $\sim \left(1 + \Delta v^2 \right)^2/\left(1-\Delta v^2 \right)$ to the overall force. For the eccentricities we are going to consider ($e\lesssim 0.2$), $\Delta v \sim 0.1$, and this is only a factor of $\sim 1-2$ (see Fig.~1 in ~\cite{Barausse:2007ph}). 

Finally, the extension of Eqs.~\eqref{eq:te_phenom}-\eqref{eq:ta_phenom} to orbits with small inclination, such that the secondary always stays immersed in the accretion disk (i.e., $e\lesssim h$) is straightforward~\cite{10.1046/j.1365-8711.2000.03466.x, 2008A&A...482..677C, 2020MNRAS.494.5666I}. Here, we focus on equatorial motion because inclination is expected to damp more efficiently than eccentricity 
as a result of the interaction with the disk~\cite{2024MNRAS.528.4958W}.



\subsection{FastEMRIWaveform package and Fisher-Information Matrix}\label{sec:FM}

With a model for the binary-disk interaction in place, we now need a framework for waveform computation and data analysis to study the detectability of gas effects with future EMRI observations. 

For the waveform, we use the modular \texttt{FastEMRIWaveform} (\texttt{few}) package~\cite{Chua:2020stf, Katz:2021yft}, which allows to input corrections to the trajectory of an EMRI by adding terms to the evolution equations governing the orbital parameters
\beq
\dot{a}_\text{m} &=& \left(\dot{a}_\text{m}\right)_\text{GW} + \left(\dot{a}_\text{m}\right)_\text{gas}   \, , \\
\dot{e} &=& \dot{e}_\text{GW} + \dot{e}_\text{gas} \, .
\eeq
Here, the subscript `GW' refers to the contribution from gravitational radiation and `gas' to the interaction with the disk, described in the previous sections. The term $\left(\dot{a}_\text{m}\right)_\text{GW}$ is the same as in vacuum GR because we neglect the disk's self-gravity (seen in Ref.~\cite{Kocsis:2008aa} to be a subdominant effect). The gravitational fluxes are computed based on self-force theory, where Einstein's equations are solved perturbatively in the mass ratio $\varepsilon$~\cite{Barack:2018yvs, Pound:2021qin}. The modified trajectory is then fed into a module to compute the waveform. A key property in this procedure is the fact the orbital phases vary on much shorter timescales than quantities that characterize the orbit, such as the orbital energy and angular momentum~\cite{Hinderer:2008dm, Miller:2020bft, Miller:2020bft}, allowing to consider orbit-averaged values for $\left(\dot{a}_\text{m}\right)_\text{gas}$ and $\dot{e}_\text{gas}$, capturing their secular evolution. For the systems we will study, typically $\left(\dot{a}_\text{m}\right)_\text{gas} / \left(\dot{a}_\text{m}\right)_\text{GW} \lesssim 10^{-3}$ and $\dot{e}_\text{gas} / \dot{e}_\text{GW} \lesssim 10^{-2}$ \textcolor{black}{in the inner disk region, and therefore even though in the supersonic regime migration is outwards ($\left(\dot{a}_\text{m}\right)_\text{gas} >0$), GW radiation-reaction dominates the evolution and the orbit shrinks, i.e. the EMRI still inspirals, although more slowly}.  

We use a fully relativistic  implementation of the Schwarzschild, eccentric waveforms publicly available in \texttt{few}
, accurate to leading order in the mass ratio $\varepsilon$ (the \textit{adiabatic} order). SMBHs are actually expected to have large spin, but a generic, fully-relativistic implementation of Kerr eccentric waveforms is currently being developed. Nonetheless, BH spin should not play a significant role in the measurement of disk effects. This is because spin corrections are more important close to plunge, while environmental effects are more relevant in the early inspiral (they are effectively a negative Post-Newtonian correction, whereas spin corrections enter at 1.5 Post-Newtonian order~\cite{poisson_will_2014}). We could use approximate, non-fully relativistic waveform models (e.g. the Augmented Analytic Kludge model~\cite{AAK_Chua_2017}), which are available for spinning BHs at larger radii, but it was recently noted that these approximants tend to underestimate the power in the early inspiral relative to the plunge~\cite{Khalvati:2024tzz}. Since this is when the interaction with the disk is more important, and the constraints change substantially when using these approximate waveforms, we opt to employ the fully relativistic ones despite their current limitations in the coverage of the parameter space.

For the data analysis, we use the Fisher-Information Matrix~\cite{Vallisneri:2007ev} to estimate the statistical uncertainty on a (single) measurement of the true source parameters. In particular, we employ the \texttt{StableEMRIFisher(SEF)} package~\footnote{ \url{https://github.com/perturber/StableEMRIFisher/tree/package} (planned to be public soon)}. The Fisher-Information Matrix is valid in the limit of high signal-to-noise ratio (SNR), assuming a flat prior in the parameter region of interest and stationary, Gaussian noise~\cite{Vallisneri:2007ev}. Under these assumptions, the posterior distribution of the source parameters $\bm{\theta}$ is given by a multivariate normal distribution around the true values, with covariance $\bm{\Sigma} = \bm{\Gamma}^{-1}$ defined by the inverse of the Fisher matrix
\begin{equation}
    \Gamma_{ij} = 4 \text{Re} \int_0^\infty \frac{1}{S_n(f)}\left(\frac{\partial \tilde{h}^*}{\partial \theta_i} \right)\left(\frac{\partial \tilde{h}}{\partial \theta_j} \right) df \, . \label{eq:Fisher}
\end{equation}
Here, $\tilde{h} = \tilde{h}(f ; \bm{\theta} )$ is a waveform template in the frequency domain (in our case computed with \texttt{few}) and $S_n(f)$ is the detector's one-sided noise spectral density~\cite{Babak:2021mhe}. More explicitly, the uncertainty in $\theta_i$ is given by the $i$'th-diagonal element of $\bm{\Sigma}$
\begin{equation}
  \sigma_{\theta_i} = \sqrt{\Sigma_{ii}}  \, . 
\end{equation}
The covariance matrix also determines the correlations between different parameters
\begin{equation}
 \langle \sigma_{\theta_i} \sigma_{\theta_j} \rangle = \Sigma_{ij}  \, . 
\end{equation}
For $S_n(f)$, we use the SciRDv1 curve for instrumental noise~\cite{scirdv}, and include the galactic confusion noise~\cite{PhysRevD.104.043019} with assumed mission duration of 4 years.

It will be useful to obtain the Fisher-Information Matrix for a new set of source parameters $\tilde{\bm{\theta}}$ having computed it from the previous set $\bm{\theta}\left(\tilde{\bm{\theta}}\right)$. Having the Jacobian  
\begin{equation}
    J_{ij} = \frac{\partial\theta_i}{\partial \tilde{\theta}_j}\Bigg|_{ \bm{\theta} = \bm{\theta}_\text{true} } \, , \label{eq:Jacobian}
\end{equation}
one just applies the transformation 
\begin{equation}
\tilde{\bm{\Gamma}}\left( \tilde{\bm{\theta}} \right) = \bm{J}^\text{T}\cdot  \bm{\Gamma} \left( \bm{\theta} \left( \tilde{\bm{\theta}} \right) \right) \cdot \bm{J} \, . \label{eq:TransformationRuleFisher}
\end{equation}

Finally, the FIM also facilitates the computation of the inference bias $\Delta \bm \theta$ in the EMRI parameters if the environmental effects are not considered in the analysis. Let us denote the disk-related parameters by $\bm \theta_\text{bv}$ (\textbf{b}eyond-\textbf{v}acuum) and the others by $\bm \theta_\text{v}$. Additionally, $\bm \theta^{0}_\text{bv}$ denotes the value of $\bm \theta_\text{bv}$ for which the environmental effects are zero, e.g. $\Sigma_0=0$.  Then~\cite{Cutler:2007mi, Kejriwal:2023djc}
\begin{equation}
    \Delta \bm\theta_\text{v} =  \left[ \left(\bm \Gamma \right )^{-1}_{\bm \theta_\text{v} \bm \theta_\text{v}} \cdot \bm \Gamma_{\bm \theta_\text{v} \bm \theta_\text{bv} } \right] \Bigg|_{\bm \theta^0_\text{bv}} \cdot \left[\bm \theta_\text{bv} - \bm \theta^{0}_\text{bv}\right] \, . 
    \label{eq:BiasFisher}
\end{equation}
In the example given above, the ambiguity in the choice of $h_0$ is irrelevant because the columns $\Gamma_{ \theta_\text{v} h_0}$ are 0. 

\section{Results}\label{sec:Results}

Having developed a model for the effect the disk has on the EMRI evolution, and a framework for the inference of the disk properties with GW observations, we now present the results of our study.

\subsection{Constraints on Accretion Physics}\label{sec:Constraints}

%
\begin{figure*}[t]
\includegraphics[width=0.315\linewidth]{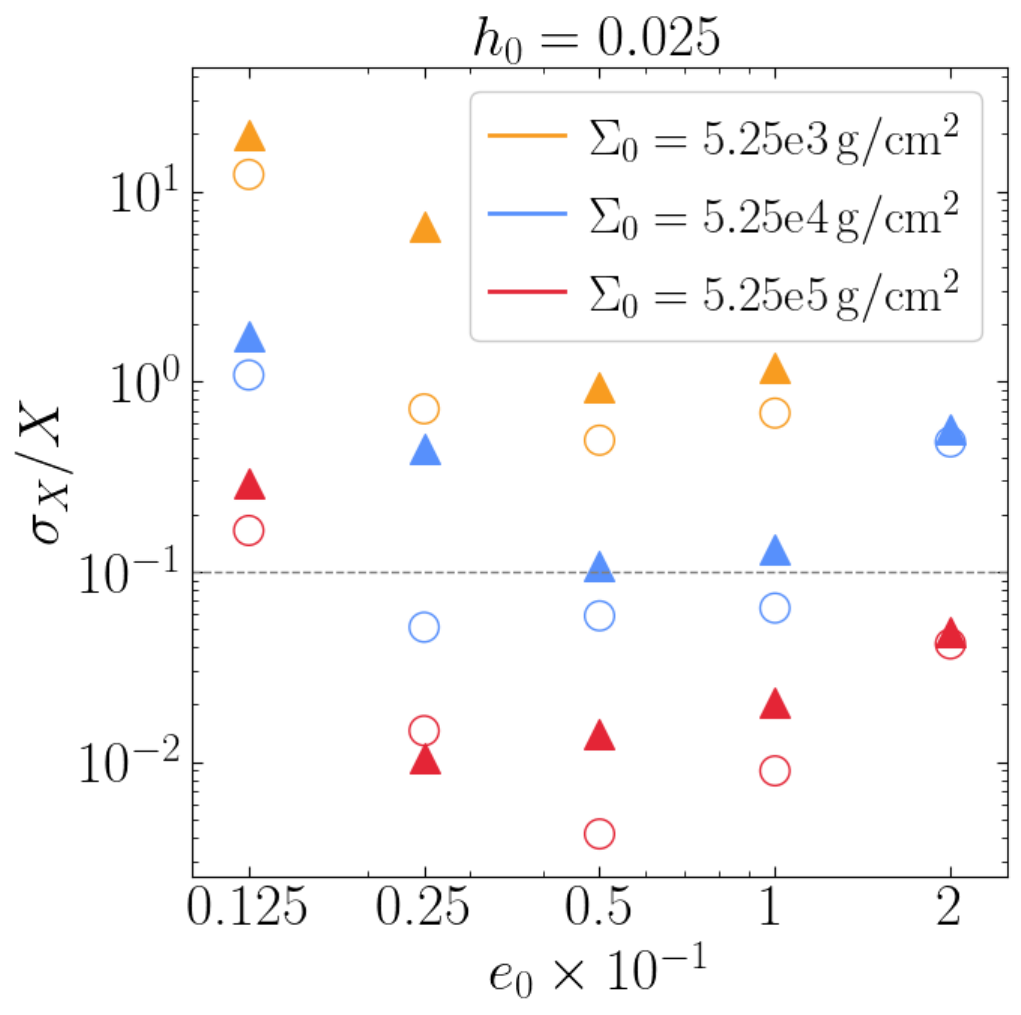}    
\includegraphics[width=0.315\linewidth]{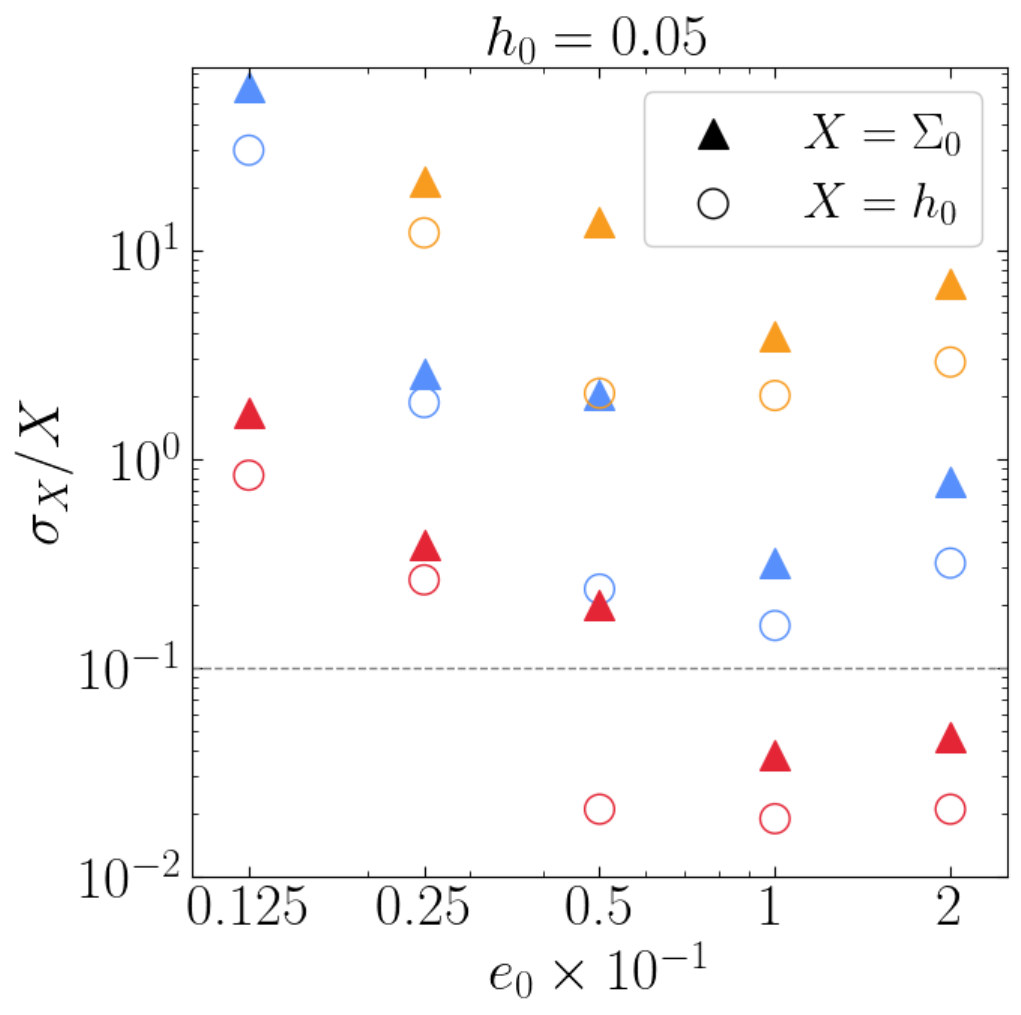}
\includegraphics[width=0.315\linewidth]{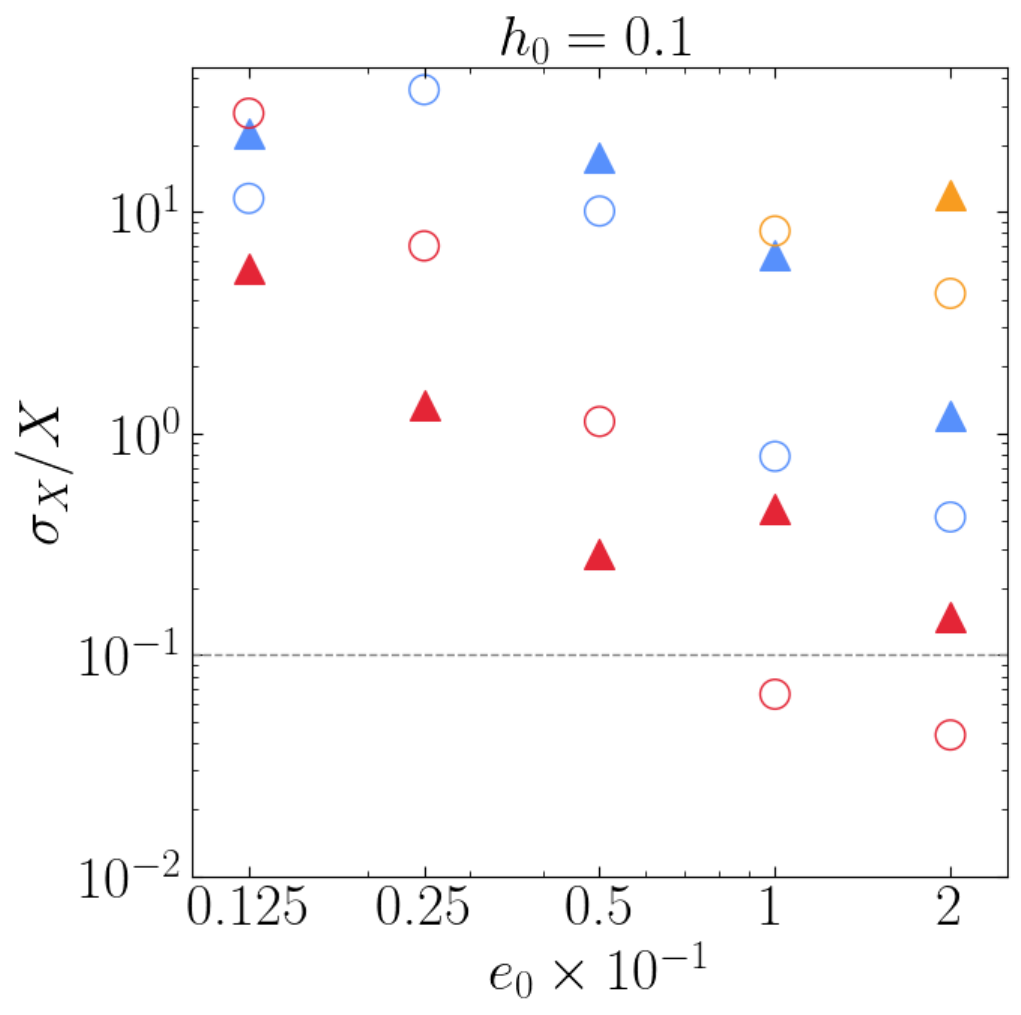}
\caption{
Measurement uncertainty computed via the Fisher-Information Matrix~\eqref{eq:Fisher} for the disk central surface density, $\Sigma_0$, and aspect ratio, $h_0$, for an EMRI evolving in different accretion configurations (different pairs $\{\Sigma_0, h_0 \}$) as a function of its initial eccentricity $e_0$. \textcolor{black}{The other parameters of the EMRI are listed in Table~\ref{tab:EMRIparameters}}. The initial semi-latus rectum is chosen so that the EMRI is close to plunge after 4 years of observation with LISA. Larger $\Sigma_0$, smaller $h_0$ and larger $e_0$ help in probing accretion physics, in particular if during the evolution the EMRI transitions from supersonic to subsonic motion relative to the gas flow in the disk, which occurs for $e \sim h$. The grey horizontal line, corresponding to $\sigma_X/X = 0.1$, serves for illustration purposes only. Given a particular disk model, constraints on $\Sigma_0$ and $h_0$ can be transformed to constraints on the accretion rate and viscosity, which we show in Fig.~\ref{fig:SigmaConversion} for the Shakura-Sunayev $\alpha$-disk model. }
\label{fig:SigmaNoConversion}
\end{figure*}
%


We focus on a ``typical'' EMRI
, \textcolor{black}{changing} only its initial eccentricity while \textcolor{black}{maintaining} the other parameters \textcolor{black}{to assess whether, for the same EMRI system, eccentricity helps in measuring the disk parameters}. The only exception is the luminosity distance, which we \textcolor{black}{change} in order to maintain the same SNR for all eccentricities. The parameters are listed in Table~\ref{tab:EMRIparameters}, and were picked to be similar to the previous study on circular inspirals~\cite{Speri:2022upm}. The initial orbital separation is chosen so that the EMRI is close to plunge after 4 years of observations. Apart from the zero spin (justified in the previous section), the EMRI's parameters are within the expected values as listed in the LISA Definition Study Report~\cite{Colpi:2024xhw}. The secondary is chosen to be heavy to mimick mass growth via accretion in the disk during EMRI formation~\cite{1944MNRAS.104..273B}. In fact, LISA could detect even heavier secondaries, in light of recent studies on intermediate-mass black holes~\cite{LIGOScientific:2020iuh, Rose:2021ftz, 2017IJMPD..2630021M}. This would yield larger SNRs and allow us to probe larger orbital separations, where the environmental effects are stronger, which would in turn improve our projected constraints on the disk parameters.
\begin{table}[t]
\label{tab:priors}
\centering
\begin{tabular}{c|c}
\toprule
\midrule
$M$ & $10^6 \, M_\odot$\\
$\mu$ & $50 \, M_\odot$\\
$a/M$ & 0 \\
$p_0/M$ & 16.83\\
$\Phi_{\varphi0}$ & $\pi/3$ \\
$\Phi_{r0}$ & $\pi/2$ \\
$\theta_S$ & 0.9\\
$\phi_S$ & 1.5\\
$T_\text{obs}$ & 4 years \\
SNR & 50 \\
\midrule
\bottomrule
\end{tabular}
\caption{EMRI parameters fixed to study the variation of the constraints on the accretion disk parameters $\{ \Sigma_0, \, h_0 \}$ as a function of the initial eccentricity $e_0$ of the EMRI (Figs.~\ref{fig:SigmaNoConversion}-\ref{fig:SigmaConversion} and Section~\ref{sec:Constraints}). The parameters are: $M$ - primary mass (detector frame); $\mu$ - secondary mass (detector frame); $a$ - primary spin; $p_0$ - initial semi-latus rectum; $\{ \Phi_{\varphi0}, \Phi_{r0}\}$ - initial phases of the periodic motion in the azimuthal and radial direction, respectively; $\{ \theta_S \, , \phi_S \}$ - polar and azimuthal sky location angles, respectively;  $T_\text{obs}$ - observation time with LISA; SNR - signal-to-noise ratio; The luminosity distance is varied for different initial eccentricities to keep the SNR fixed. For the parameters listed the luminosity distance is approximately $d_L \sim 4 \, \text{Gpc}$. } \label{tab:EMRIparameters}
\end{table}

We pick a set of 5 initial eccentricities -- $e_0 = \{0.0125, \,  0.025, \, 0.05, \, 0.1, \, 0.2   \}$ -- and compute the Fisher-Information Matrix~\eqref{eq:Fisher} for \textcolor{black}{the full parameter space of} the EMRI system in Table~\ref{tab:EMRIparameters} and $9$ different disk configurations, corresponding to different pairs of the central surface density ---  $\Sigma_0 = 5.25 \times \{10^3, \,  10^4, \, 10^5  \} \, \left[\text{g}/\text{cm}^2 \right]$ --- and the central aspect ratio --- $h_0 = \{0.025, \,  0.05, \, 0.1\}$. As mentioned when the disk model was introduced in Sec.~\ref{sec:TorqueModel}, these are compatible with numerical simulations and observations.  

In Fig.~\ref{fig:SigmaNoConversion} we show the relative error computed from the Fisher-Matrix for $\Sigma_0$ and $h_0$ for all configurations. The statistical errors for the other EMRI parameters are of the same order of magnitude as in vacuum. We thus choose to only show the full corner plots for two configurations in Appendix~\ref{app:CornerFisher}, one corresponding to initial subsonic motion and the other to initial supersonic motion. 

Larger densities and smaller aspect ratios lead to larger corrections to the inspiral, independently of whether the EMRI is in the subsonic or supersonic regime, and thus these configurations optimize the constraints on the disk parameters. For $h_0 = 0.1$, it is not possible to probe the effect of the accretion disk for any of the eccentricities, since all measurements of the density would be consistent with 0, i.e., no disk, within $1-\sigma$ uncertainty. This is also true for all configurations with $\Sigma_0 = 5.25 \times 10^3 \, \text{g}/\text{cm}^2 $. To probe these densities/aspect ratios would require EMRIs with larger SNR, e.g., closer to Earth, and/or that start from larger orbital separations, e.g., with larger mass ratio. 

Nonetheless, there are a couple of configurations for which both $\Sigma_0$ and $h_0$ would be measurable with a relative error smaller than $10 \, \%$, even with $e_0$ as small as $0.025$ ($h_0 = 0.025$ in the left panel). We note that this is not possible for circular motion, for which these parameters only enter the correction to the semi-major axis through the ratio $\Sigma_0/h_0^2$, and are therefore completely degenerate. It is also clear that eccentricity plays a major role in `deviating' the wet EMRI from a typical vacuum inspiral. In fact, throughout the inspiral the relative difference in the eccentricity evolution between in-disk and in-vacuum is typically one order of magnitude larger than the same relative difference in the semi-latus rectum evolution. 

The initial eccentricity that minimizes the measurement uncertainty varies with $h_0$, and is related to whether the EMRI transitions from the super to the subsonic regime during the inspiral. As seen earlier, during this transition the functional dependence on $h_0$ of the rate of change of the orbital parameters changes (check Table~\ref{tab:Sub_vs_Super_Timescales} and Fig.~\ref{fig:te_ta}). Although this complicates the model, it likely breaks correlations with other parameters. When the initial eccentricity is either substantially smaller or larger than the aspect ratio, the EMRI spends most of its inspiral in, respectively, the subsonic or supersonic regime. This effectively makes the correction to the trajectory due to the environment simpler and, therefore, harder to constrain, as seen in the panel for $h_0=0.025$. Also recall that in the transition from super-to-subsonic motion, there is an enhancement in the damping of the orbital separation due to the resonant excitation of sound waves in the gas (the bump at $e\sim 2h$ in the right panel of Fig.~\ref{fig:te_ta}). In Appendix~\ref{app:Matching}, we show results for a smoother profile (see Fig.~\ref{fig:ta_comp}), and conclude the qualitative behavior observed in Fig.~\ref{fig:SigmaNoConversion} is maintained. In fact, a bump less pronounced does not necessarily imply measurements less precise, as illustrated in Fig~\ref{fig:ConstraintsComp}. 

Despite the measurement uncertainties being of the same order of magnitude as in vacuum, we observe a change in the nature of some correlations depending on whether the EMRI starts in subsonic or supersonic motion. For example, in vacuum, there is typically a strong negative correlation between $M$ and $p_0$, which is also observed in the fully subsonic case (Fig.~\ref{fig:FullCorner_Subsonic}). Intuitively, as the initial semi-latus rectum increases, so does the time to plunge, which can be compensated by a decrease in the primary mass. However, as the initial ratio $e_0/h_0$ increases, indicating the EMRI starts the inspiral in the supersonic regime, the $M-p_0$ correlation becomes positive (Fig.~\ref{fig:FullCorner_Supersonic}), which was confirmed for different $h_0$. Similar sign changes occur, for example, for $M-e_0$ and $p_0-\Sigma_0$ correlations. We attribute this to the intricate transonic behavior of the inspiral. In fact, if the EMRI's initial eccentricity is very large (e.g., $e_0 = 0.6$) and it spends the whole observation window in the supersonic regime, we recover the same correlations as in the subsonic case. We note that changes in the sign of correlations have been observed in other EMRI systems, e.g., eccentric EMRIs with scalar charge~\cite{Speri:2024qak}, or when trying to recover a vacuum signal with an approximate template~\cite{Katz:2021yft}. We also confirmed the stability of the Fisher-Matrix in these cases by adopting different parameterizations, in particular by working in $\log M$ and $\log \mu$, and checked that the $M-p_0$ correlation changes smoothly as $e_0$ increases, by computing more values of $e_0$ close to each other. 

Similarly to circular EMRIs~\cite{Speri:2022upm}, we found for some disk configurations significant bias in the intrinsic parameters,  shifted $3\sigma$ away from their true value, if the effect of the disk is not taken into account in the inference. Conversely, the extrinsic parameters (sky location and luminosity distance) displayed much smaller bias. Yet, we found the Fisher-Matrix was more prone to numerical instabilities when computed at the null hypothesis, i.e. at the values of the disks parameters $\bm \theta = \bm \theta^0_\text{bv}$ for which their effect is null, which is necessary for the bias computation according to Eq.~\eqref{eq:BiasFisher}. We therefore leave for future work a more thorough study of these biases over the parameter space. As shown in Ref.~\cite{Speri:2022upm}, the bias can be so significant after 4 years of observation that the signal can be missed in a search of the full data set. The signal could still be caught using smaller chunks of data, but one would see the maximum likelihood estimators drifting from the true value as the observation time increases (and SNR not accumulating). This would be a data-driven smoking gun of the presence of a beyond-vacuum effect.

If we specify a particular accretion disk model, we can relate the surface density and aspect ratio with internal disk properties like accretion rate and viscosity, as made explicit in Eqs.~\eqref{eq:Shakura-Sunayev_Sigma}-\eqref{eq:Shakura-Sunayev_h} for the Shakura-Sunayev $\alpha$-disk. Then, using Eqs.~\eqref{eq:Jacobian}-\eqref{eq:TransformationRuleFisher}, we can obtain the Fisher-Information matrix for this new set of parameters and compute their measurement uncertainty.  We show this in Fig.~\ref{fig:SigmaConversion} for $h_0=0.025$. For some systems one can simultaneously constraint the viscosity, $\alpha$, and the accretion rate-efficiency ratio, $f_\text{Edd}/\epsilon$. As mentioned before, this is not possible in circular orbits where these parameters become completely degenerate and one can only constrain an overall amplitude in $(\dot{a}_\text{m})_\text{gas}$ which captures some combination of $\alpha$ and $f_\text{Edd}/\epsilon$ (check Eq.~(2.2) in~\cite{Speri:2022upm}). To then break this degeneracy, one would need an electromagnetic counterpart to the GW signal to yield an independent measurement of the accretion rate. 





\subsection{Formation Channels}\label{sec:Formation}

The differences between the supersonic and subsonic interactions also have implications 
for the formation of wet EMRIs. As commented in the Introduction, previous studies on these systems argued they should be (\textit{quasi})circular~\cite{Pan:2021oob, Pan:2021ksp} because, \textit{in the subsonic regime}, density waves damp eccentricity on a much faster timescale than migration. One can see this explicitly from the first row in Table~\ref{tab:Sub_vs_Super_Timescales} where $t_e/t_{a_\text{m}} \sim h^2 \ll 1$. However, as seen above, \textit{in the supersonic regime}, local dynamical friction is the dominant environmental effect \textcolor{black}{which alters the functional form for the evolution of the orbital parameters and impacts the ratio $t_e/t_{a_\text{m}}$. We focused our analysis in the inner disk region ($r\lesssim 100M$), where GW radiation reaction dominates over environmental effects. Therefore, EMRIs that enter the inner disk region with eccentricity large enough so that \textit{in vacuum} they would enter the LISA band with significant eccentricity, should also do so in the presence of the disk. Recall that LISA can measure the eccentricity of EMRIs at coalescence with absolute precision of $10^{-5}$ and that EMRIs formed via multi-body scattering in vacuum are expected to enter the LISA band with $e_0 \lesssim 0.7$~\cite{2017PhRvD..95j3012B, Colpi:2024xhw}.} 

\textcolor{black}{Can a wet EMRI evolving from larger radii in the disk sustain significant eccentricity up to when it enters the inner disk region? To answer this, we need to extend the model in Eqs.~\eqref{eq:te_phenom}-\eqref{eq:ta_phenom} to the outer disk region ($r \gtrsim 100M$), where $\Sigma_p = 1/2$ such that $\Sigma \sim \Sigma_0 /r^{1/2}$ and $h \sim \text{constant}$ (valid up to $r \lesssim 10^4M$ for $M \sim 10^{6}\, M_\odot$~\cite{Gangardt:2024bic}). We present this extension in Appendix~\ref{app:OuterDisk}, with the analytic estimates in both limits already listed in Table~\ref{tab:Sub_vs_Super_Timescales}. First, we find that migration is always \textit{inwards} ($\dot{a}_\text{m} <0$), since the density profile decays with radius and therefore, contrarily to the inner disk, the ''headwind'' at the periapsis dominates over the ``tailwind'' at the apoapsis, decelerating the binary (recall the discussion below Eq.~\eqref{eq:arelpe}). Moreover, in the subsonic regime, the ratio $t_e/t_{a_m} \sim h^2 $, as in the inner disk region. However, in the supersonic case, we now have ${t_e/t_{a_m} \sim 0.6 e^2/(1-e^2)^{3/4}}$, which is larger than 1 for $e \gtrsim 0.85$ and can become very large for $e \sim 1$ (note the change in the dependence of $t_e$ and $t_a$ with $(1-e^2)$ for different disk models had already been reported in Ref.~\cite{2011ApJ...737...37M}). Thus, eccentricity damping can occur on timescales much larger than migration in the outer disk region and at large eccentricities, reducing the circularizing effect of migration.}  

\textcolor{black}{Let us consider the case of the fiducial EMRI listed in Table~\ref{tab:EMRIparameters}, with $e_0 = 0.1$ and disk configuration with $\Sigma_0 = 5.25 \times 10^5 \, \text{g}/\text{cm}^2$ and $h_0 = 0.05$. If we integrate its trajectory backwards, including GW radiation reaction and gas effects, combining both the inner and outer disk models (with transition between them placed at $r \sim 100M$), we find that $e = 0.9$ at $a_\text{m} \sim 220 M$. An EMRI with this initial conditions in vacuum would then reach the same semi-latus rectum  as quoted in Table~\ref{tab:EMRIparameters} ($p = 16.83M$), with eccentricity $e_0 \sim 0.25$, so within the expected range for vacuum EMRIs in LISA, which should even be captured inclined and at larger radii}. This example puts into question the expectation that wet EMRIs will always be circular. There are also other environmental interactions which excite eccentricity, such as stochastic fluctuations in the gas flow~\cite{2005A&A...443.1067N, Johnson:2006eh},  retrograde orbits~\cite{Sánchez-Salcedo_2018, Sánchez-Salcedo_2020}, or dynamical friction in dark matter distributions~\cite{Cardoso:2020iji, Tomaselli:2024bdd}.  

%

Recently, Ref.~\cite{2024MNRAS.528.4958W} considered the impact of some of these effects in the formation of stellar-origin BH binaries in AGN disks, concluding that for initial retrograde orbits `there might be some residual eccentricity in captured BHs'. It is clear from the examples above that the assumption of quasi-circularity for EMRIs forming in accretion disks needs to be \textcolor{black}{evaluated more carefully}. We leave for future work a quantitative study of the effect that stochasticity and interaction with a DM distribution have on the eccentricity distribution of a population of wet EMRIs. We also note that equal mass binaries surrounded by circumbinary disks have been recently found to evolve to fixed points in eccentricity~\cite{10.1093/mnras/stae569, Franchini:2024pgl} (see also Ref.~\cite{Tomaselli:2024bdd} for a similar behavior in ultralight dark matter environments).  

Reference~\cite{2024MNRAS.528.4958W} also found inclination is always zero (for the purposes of LISA precision) by the time the EMRI is captured and eventually enters the LISA band, independently of its initial conditions. Nonetheless, inclined orbits are relevant in the context of periodic electromagnetic signals associated to EMRIs at large radial separations, which we discuss in more detail in the next subsection. 

\subsection{Host identification and multimessenger EMRIs}~\label{sec:Multimessenger}

%
\begin{figure}[t]
\centering
\includegraphics[width=0.85\linewidth]{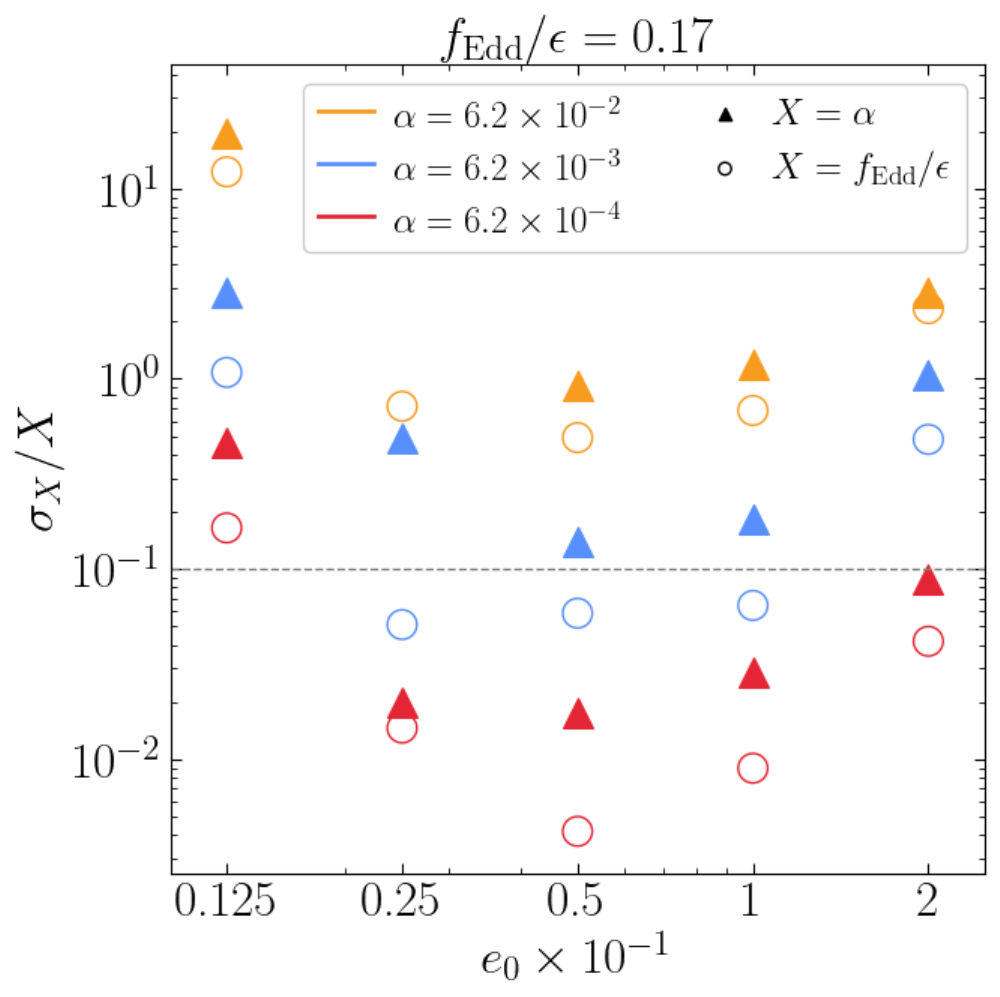}    
\caption{Same as in the left panel of Fig.~\ref{fig:SigmaNoConversion} ($h_0 = 0.025$), but transforming constraints in the surface density and aspect ratio $\{\Sigma_0, \, h_0 \}$ into constraints on the accretion rate and viscosity $\{f_\text{Edd}/\epsilon, \, \alpha \}$ for the Shakura-Sunayev model~\eqref{eq:Shakura-Sunayev_Sigma}-\eqref{eq:Shakura-Sunayev_h}. 
}
\label{fig:SigmaConversion}
\end{figure}

We now discuss the possibility of detecting an electromagnetic (EM) counterpart to eccentric wet EMRIs, making these sources candidates for standard sirens. This can be accomplished by identifying the host galactic nucleus, and/or by observing an EM counterpart directly linked to the EMRI's orbital motion.

Let us start with the identification of the host galaxy. 
By the time LISA flies, the Vera C.~Rubin Observatory Legacy Survey of Space and Time (LSST) will have provided the most complete AGN catalog, down to the luminosities of Seyfert galaxies\footnote{Seyfert galaxies are a class of AGNs with moderate luminosity ($10^{42} - 10^{44} \, \text{erg/s}$) in comparison to Quasars ($10^{45}-10^{48} \, \text{erg/s}$)} and up to high redshift, and covering $18000\, {\rm deg}^2$  -- almost half of the sky \cite{LSST_ScienceBook2009}. The LSST catalog will have redshift errors of order $\Delta z \sim \pm 0.05$  at low redshifts  \cite{LSST_ScienceBook2009}, which are sufficiently small compared to the ones obtained for EMRI events, $\Delta z \sim \pm 0.06$ (assuming some cosmology). 

Suppose we then detect an EMRI with significant evidence for an environmental effect described by the model we studied. From the GW signal we will measure, with high precision for astrophysical purposes, the SMBH mass and spin (e.g. $\sigma_{M_S} / M_S \sim 10^{-2} , \,  \sigma_{a} / a \sim 10^{-5}-10^{-4}$, where $S$ denotes source frame mass and $a$ the BH spin)~\cite{Colpi:2024xhw}, see also Fig.~\ref{fig:FullCorner_Subsonic}-\ref{fig:FullCorner_Supersonic}), and also the ratio $f_\text{Edd}/\epsilon$  (Fig.~\ref{fig:SigmaConversion}). From the spin, we can estimate the mass-radiation conversion efficiency $\epsilon$~\cite{Madau:2014pta, Mangiagli:2022niy}, from which we can then infer the luminosity Eddington ratio $f_\text{Edd}$. With the latter we compute the bolometric luminosity of the AGN hosting the EMRI through $\dot{L} = f_\text{Edd} L_\text{Edd}$, where $L_\text{Edd}=1.26 \times10^{38}\left(M/M_\odot \right) \, \text{erg/s} = 3.2\times 10^4 \left(M/M_\odot \right) L_\odot$, with $L_\odot = 3.846 \times 10^{33}\, \text{erg/s}$ being the Sun's luminosity. The GW signal also gives a precise measurement of the luminosity distance $d_L$. By combining $\dot{L}$ and $d_L$, we can infer the \textit{apparent} magnitude\footnote{The magnitude measures the brightness of a source in an inverse logarithmic scale, such that brighter objects have smaller magnitude.} of the AGN in the optical band of the Rubin observatory through~\cite{Mangiagli:2022niy} 
\begin{equation}
    m_\text{AGN, Rubin} = M_\text{AGN, Rubin} + 5 \, \log_{10} \left(\frac{d_L}{\text{pc}} \right) - 5 \, ,  
\end{equation}
where $M_\text{AGN Rubin}$ is the \textit{absolute} magnitude (independent of distance) in that instrument's band~\cite{Mangiagli:2022niy} 
\begin{equation}
    M_\text{AGN, Rubin} = 4.64 - 2.5\log_{10} \left(\frac{0.1L}{{L_\odot}} \right)\, .  
\end{equation}
The Rubin Observatory will be sensitive to $m_\text{AGN, Rubin} \lesssim 27.5$ (see Section IV in Ref.~\cite{Mangiagli:2022niy} for generic expressions for other frequencies/instruments). 

Assuming a fiducial value of $\epsilon = 0.1$, for the systems studied in Section~\ref{sec:Constraints}, only disk configurations with $h_0 = 0.1$ ($f_\text{Edd}/\epsilon = 0.068$ for the $\alpha$-disk) would yield an apparent magnitude $m_\text{AGN, Rubin} = 26.8$, smaller than the Rubin threshold. However, for the same EMRI at half the luminosity distance ($d_L \sim 2 \, \text{Gpc}$), all disk configurations would have detectable magnitudes in the catalog. Higher primary masses and conversion efficiencies, which can go up to $\epsilon \sim 0.4$ for maximally rotating BHs, would also decrease the AGN's magnitude~\cite{Madau:2014pta, Mangiagli:2022niy}. Taking the measurement uncertainties for these brighter sources to be similar to the ones obtained in Sec.~\ref{sec:Constraints} for the most constrained systems ($\Delta f_\text{Edd} / f_\text{Edd} \sim 0.1$), we find that the GW signal alone gives an estimate of the magnitude with relative error $\Delta m_\text{AGN, Rubin} / m_\text{AGN, Rubin} \sim 0.01$, whereas the photometric errors by Vera Rubin are $0.005-0.02$, depending on brightness. We could then compare within the EMRI's sky location error ($\Delta \Omega \sim 2-18 \, \text{deg}^2$~\cite{Colpi:2024xhw}), how many AGNs in the LSST have magnitudes matching the GW estimate. 
Even if multiple AGNs were compatible, one could do more targeted searches in the sky area identified by the GW measurement, either with Vera Rubin, or with instruments like the Square Kilometer Array (radio)~\cite{5136190}, or ATHENA (X-Ray band)~\cite{2013arXiv1306.2307N}. Having a reduced number of possible host galaxies for the EMRI could also improve their use as dark sirens~\cite{Laghi:2021pqk}.

\section{Conclusions \& Discussion}~\label{sec:Conclusions}

In this work, we developed a model for the evolution of eccentric EMRIs in accretion disks valid up to \textcolor{black}{large} eccentricities. Our model connects analytic results from Newtonian analysis when the small compact object is moving subsonically or supersonically with respect to the gas flow of the disk. While in the subsonic limit the EMRI excites spiral density waves that resonantly exchange angular momentum with the binary throughout the whole disk (\textit{migration}), in the supersonic case the dominant contribution comes from the local density wake that trails behind the secondary (\textit{dynamical friction}). These two regimes are ``glued'' together with hydrodynamical simulations capturing the transition between then. 

Using our model, we showed that eccentricities as small as $e \sim 0.025$ when the binary enters the LISA band can significantly improve the measurement of the environmental effects caused by the binary-disk interaction and consequently the inference of the disk properties, like its density/height or viscosity/accretion rate. Their simultaneous constraint is not possible for circular orbits, where they become fully degenerate, and for which EM identification is needed to measure both viscosity and accretion rate. The fact that for eccentric wet EMRIs we can obtain an estimate for the primary's mass accretion rate solely from the GW signal offers the possibility of identifying the host galaxy of the system in an AGN catalog, like the LSST survey by the Vera C. Rubin Observatory. 

Direct EM counterparts to the GW event could also be observed with targeted searches for periodic variability in luminosity consistent with the EMRI's orbital period, like the recently detected Quasi-Periodic Eruptions/Oscillations(QPE/O)~\cite{Kejriwal:2024bna, 2023ApJ...957...34L, 2023A&A...675A.100F, Arcodia:2024taw, 2024SciA...10J8898P}. \textcolor{black}{These are periodic soft X-Ray signals (\textcolor{black}{with typical disk temperatures} $\sim 0.1\, \text{keV}$) thought to originate from a compact object (secondary) passing roughly twice per orbit through an accretion disk around a SMBH. The orbit is necessarily inclined and several models have been proposed, involving, e.g., the tidal disruption of a main-sequence star~\cite{2023ApJ...957...34L}, or disk precession~\cite{2023A&A...675A.100F}, \textcolor{black}{all within the EMRI modeling framework, where the secondary-disk interactions affect the binary's dynamics and lead to the periodic variability of the QPE/O X-ray emission peaks. While alternate models can account for a subset of such soft X-ray band low-frequency QPE/Os~\cite{Ingram:2021gar,Miniutti:2019fqr}, the highly dynamic EMRI model provides a broader scope. Many QPE/Os have been identified in galaxies with a SMBH with mass $\sim 10^6 \, M_\odot$~\cite{10.1093/mnrasl/slad113}, precisely in the mass range interesting for LISA EMRIs. However, their frequencies are typically $\lesssim 10^{-4} \, \text{Hz}$, indicating that the secondaries are \textit{currently} at large orbital separation. 
EMRIs emitting today in the soft X-ray band as QPE/Os with frequencies $\approx 0.5$ mHz (periodicity $\sim$ hours) may evolve to LISA's GW sensitivity band in $\approx 10$ years when it will start operation, which can enable multi-messenger EMRI inference~\cite{Kejriwal:2023djc}.
As EM observatories improve, 
it may also be possible to observe} QPE/Os closer to the central SMBH and directly search for X-Ray variability in the EMRI's sky location after its detection. Yet, to better understand the time evolution of the EMRI's orbit and its corresponding EM counterpart, it is crucial to study the transition from out-of-equatorial plane to in-plane motion.}

Despite our simplistic model, drawn mostly from works on (Newtonian) planetary migration, and the limited region of the parameter space we explored, our results highlight the importance of more detailed modelling of beyond-vacuum GR effects in binary coalescence for future GW astronomy. Just a simple refinement from previous studies of EMRIs in accretion flows -- the inclusion of eccentricity -- leads to more interesting phenomenology which can act as distinctive smoking-guns for these astrophysical environments, enhancing their detectability. This is particularly relevant considering the multiple degeneracies in EMRI parameter space~\cite{Chua:2021aah}. These can potentially hinder constraints on beyond-vacuum GR corrections, especially when multiple ones are considered simultaneously in the analysis of the GW signal~\cite{Kejriwal:2023djc}. As better models are developed, we will also be able to assess their relevance when compared to the precision of future GW detectors, i.e., how large systematics in the modelling of the environmental effects are compared to the statistical uncertainty in their measurement. 

Our study should by no means be taken as a final answer to this problem, but rather as yet another example of the rich physics associated to environmental effects in the formation and evolution of compact binaries. We stress that there is room for improvement in our model, by considering more sophisticated disk strucutures (e.g., numerical profiles), stochastic oscillations in the gas flow, inclined orbits, mass accretion onto the secondary and relativistic corrections. An important follow-up would be to compare our results for eccentric inspirals with state-of-the-art hydro codes capable of evolving I/EMRIs immersed in accretion disks. This comparison has been done only for circular orbits~\cite{Derdzinski:2020wlw}, where it was found that for sufficiently light secondaries and ``warm'' disks (large viscosities), the overall effect in the binary is well captured by the analytic migration formula, up to some numerical factor of $\mathcal O(1)$. However, when the disk relaxation timescale is comparable to the binary's inspiral timescale, which occurs for heavier secondaries and colder disks (smaller viscosities), the binary-disk interaction becomes chaotic and the forces felt by the secondary exhibit strong fluctuations, differing from the analytic result. In this limit, the orbital evolution is no longer adiabatic, i.e., we can no longer consider orbit-averaged effects. In fact, this brings us to the more generic problem of how to accurately include environmental effects in the framework of self-force theory, which relies on the two-timescale expansion to separate the orbital timescale, $T_\text{o} \sim M$, from the inspiral timescale, $T_\text{i} \sim M^2/\mu$, on which quantities that characterize the orbit vary (e.g., energy, angular momentum).   


Our analysis of measurability of these effects with LISA was also simplistic. Complementing the Fisher-Matrix with fully Bayesian inference using Markov Chain Monte Carlo methods would be valuable. Even more relevant would be understanding how to search for deviations from vacuum in the context of the LISA Global Fit~\cite{Littenberg:2023xpl, Katz:2024oqg, Strub:2024kbe}. LISA will observe $\sim 10^4$ different signals --  generated by galactic binaries, EMRIs, massive binaries, stochastic GW backgrounds and other sources -- which need to be searched and analysed simultaneously~\cite{Speri:2022kaq}. One of the hardest challenges of this scheme are precisely EMRIs, which even in vacuum exhibit a very degenerate parameter space, with numerous secondary modes~\cite{Chua:2021aah}. We are then faced with the question of how to include beyond-vacuum-GR effects on top of the vacuum problem. Doing it directly in the Global Fit pipeline would be technically demanding and could become too computationally expensive. However, since these effects are typically small corrections (as it is the case here), we can envision a scenario where we use the residuals obtained from the vacuum-only Global Fit to reanalyse a single source (e.g., the EMRI with largest SNR), to search for additional physics, like the disk interaction considered in this work. Further work is needed to understand if such an approach would work and be free from systematics. We plan to address some of the problems listed above in the near future.

\acknowledgments
We thank the anonymous referee for the thoughtful comments that helped to substantially improve the original version of this manuscript. We thank B.~McKernan, A.~Dittmann, N.~Tamanini, D. Laghi and A. Mangialgi for helpful discussions on the identification of the host and potential electromagnetic counterparts. We also thank R.~Vicente for fruitful insights on dynamical friction in gaseous media and Vojt\v{e}ch Witzany for comments on the effect of turbulence to migration. We also thank A.~Santini for support in plotting through his \texttt{pysco} package~\cite{pysco_2024_13930440}. This work used NVIDIA A100 GPUs through the ``Saraswati'' facility at the Max Planck Institute for Gravitational Physics. 

\bibliography{references}

\appendix

\newpage

%

\section{Torque model for the outer disk}\label{app:OuterDisk}

In this Appendix, we extend to the outer disk region ($r \gtrsim 100M$), the phenomenological model valid to large eccentricities for the evolution of the EMRI's orbital parameters due to the interaction with gas. In this disk region, the power-law model for the density and aspect ratio profile in Eqs.~\eqref{eq:PowerLawSurface}-\eqref{eq:PowerLawAspectRatio} takes $\Sigma_p = 1/2$, such that $\Sigma = \Sigma (10M/r)^{1/2}$ and $h = \text{constant}$. Repeating the same steps as in Sec.~\ref{sec:TorqueModel}, we find the timescales for the evolution of orbital parameters in the supersonic regime in this disk region become
\beq
&&\frac{1}{t^\text{sup}_{a_\text{m}}} \propto \frac{\Sigma_0}{h_0}\frac{(1-e^2)^{1/4}}{e}  \propto  \frac{1}{t_\text{gas}} \frac{h^3}{e}\left(1 - e^2\right)^{1/4}  \, , \label{eq:ta_sup_outer} \\
&&\frac{1}{t^\text{sup}_e} \propto \frac
{\Sigma_0}{h_0}\frac{(1-e^2)}{e^3} \propto \frac{1}{t_\text{gas}}\left(\frac{h}{e}\right)^3 \left(1 - e^2\right) \, . \label{eq:te_sup_outer}
\eeq
The proportionality factors are presented in Table~\ref{tab:Sub_vs_Super_Timescales} in the entry for the outer disk. 

Two comments are in order. Firstly, in this disk region dynamical friction leads to inward migration, contrarily to what occurred for the inner disk. This is related to the decay of the surface density with radius, which makes the the ''headwind'' at the periapsis dominate over the ``tailwind'' at the apoapsis, decelerating the binary (recall the discussion below Eq.~\eqref{eq:arelpe}). This behavior is important to guarantee the EMRI actually inspirals, since for sufficiently large radius in this disk region, the interaction with the gas dominates over GW radiation reaction for the orbital evolution (the radius where this occurs is dependent on the disk central density, scale height and eccentricity, but typically is $200M \lesssim r \lesssim 1000M $). Secondly, $t^\text{sup}_{a_\text{m}}$ and $t^\text{sup}_{e}$ show different dependences in the quantity $(1-e^2)$ compared to the inner disk region (compare the above with Eqs.~\eqref{eq:ta_sup}-\eqref{eq:te_sup}), in particular now $t^\text{sup}_{e}$ grows faster than $t^\text{sup}_{a_\text{m}}$ as $e \rightarrow 1$. This has implications on the possibility of forming eccentric wet EMRIs, as discussed in detail in Sec.~\ref{sec:Formation}. We note that the change in the dependence of $t_e$ and $t_a$ with $(1-e^2)$ for different disk models had already been observed in previous studies~\cite{2011ApJ...737...37M}).

\begin{figure}[h]
\includegraphics[width=0.99\linewidth]{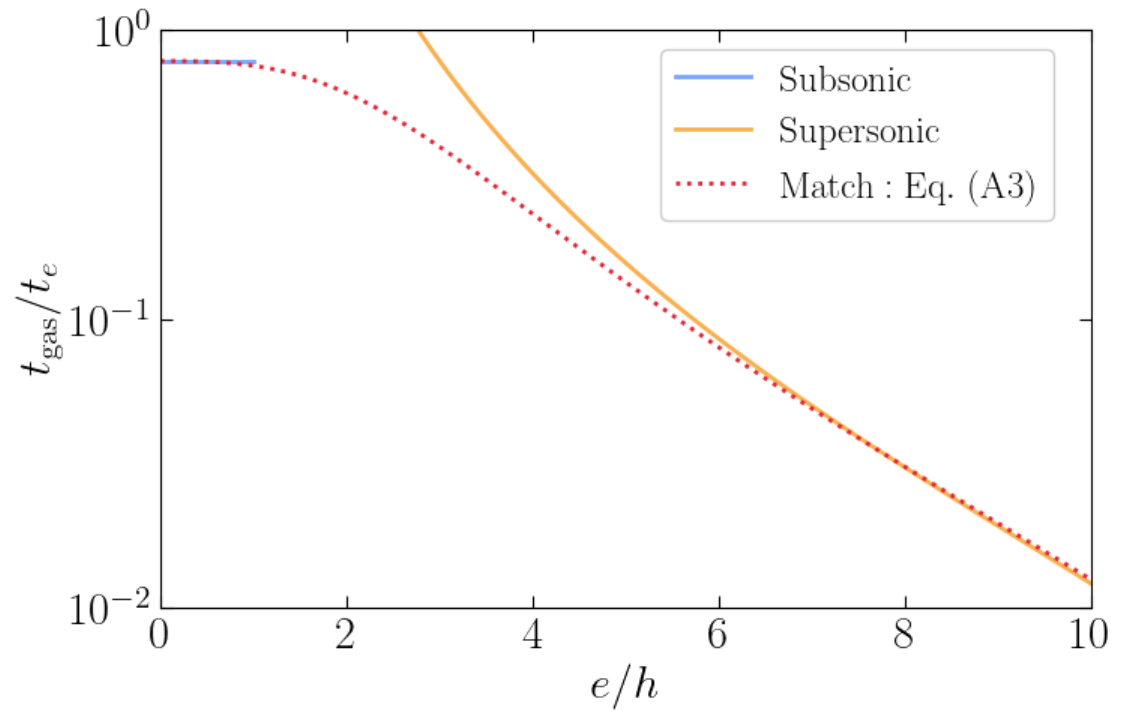}  
\includegraphics[width=0.99\linewidth]{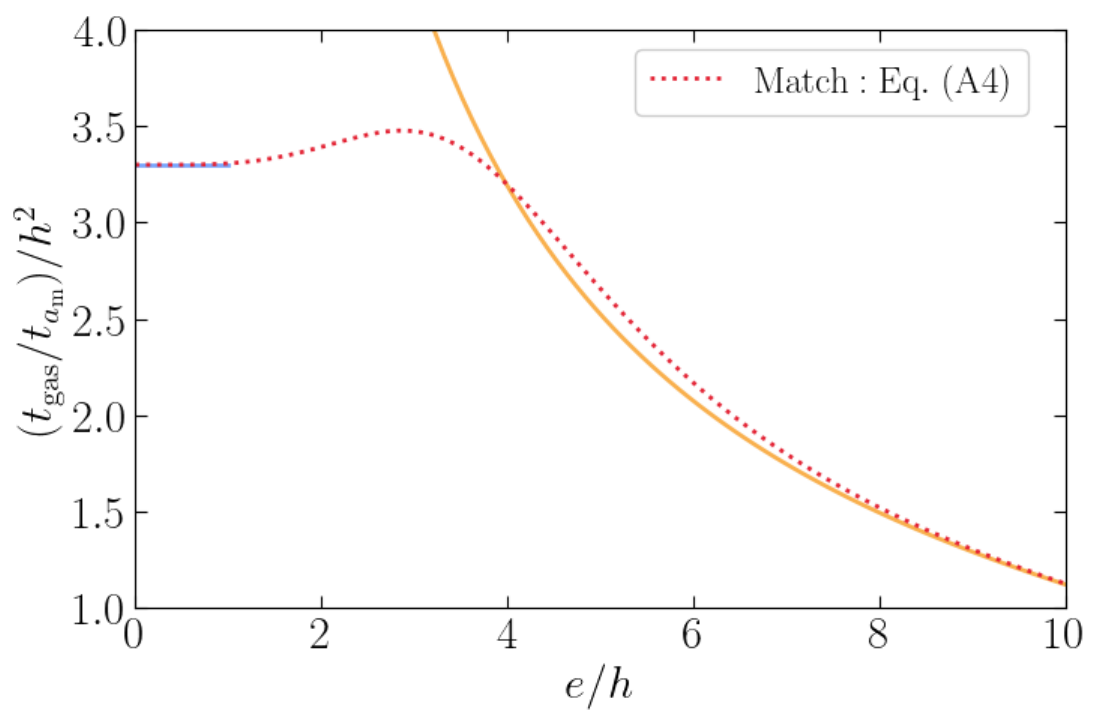}     
\caption{Same as in Fig.~\ref{fig:te_ta} but for the outer disk region ($r \gtrsim 100M$), where $\Sigma_p = 1/2$ in the density and aspect ratio profiles in Eqs.~\eqref{eq:PowerLawSurface}-\eqref{eq:PowerLawAspectRatio}.}
\label{fig:timescales_outer}
\end{figure}

Combining the results above with the linear estimates for migration in the subsonic regime, we propose the following phenomenological model for the outer disk region
\beq
\frac{t_\text{gas}}{t_e} &=& 0.78\left(1 - e^2\right)\Big/\left(1+\frac{1}{30}\left(\frac{e}{h}\right)^3 \right) ,  \label{eq:te_phenom_outer} \\
\frac{t_\text{gas}}{t_{a_\text{m}}} &=& 2 C_\text{Sub}\, h^2 \left(1-e^2\right)^{1/4}\frac{1+\left(\frac{e}{3}\right)^4}{1+\left(\frac{e}{3h}\right)^5} \, . \label{eq:ta_phenom_outer}
\eeq
which is illustrated in Fig.~\ref{fig:timescales_outer}, compared to the analytical estimates given in Table~\ref{tab:Sub_vs_Super_Timescales} for the subsonic and supersonic limits.

\section{Different Migration Models}\label{app:Matching}

\begin{figure}[t]
\includegraphics[width=0.99\linewidth]{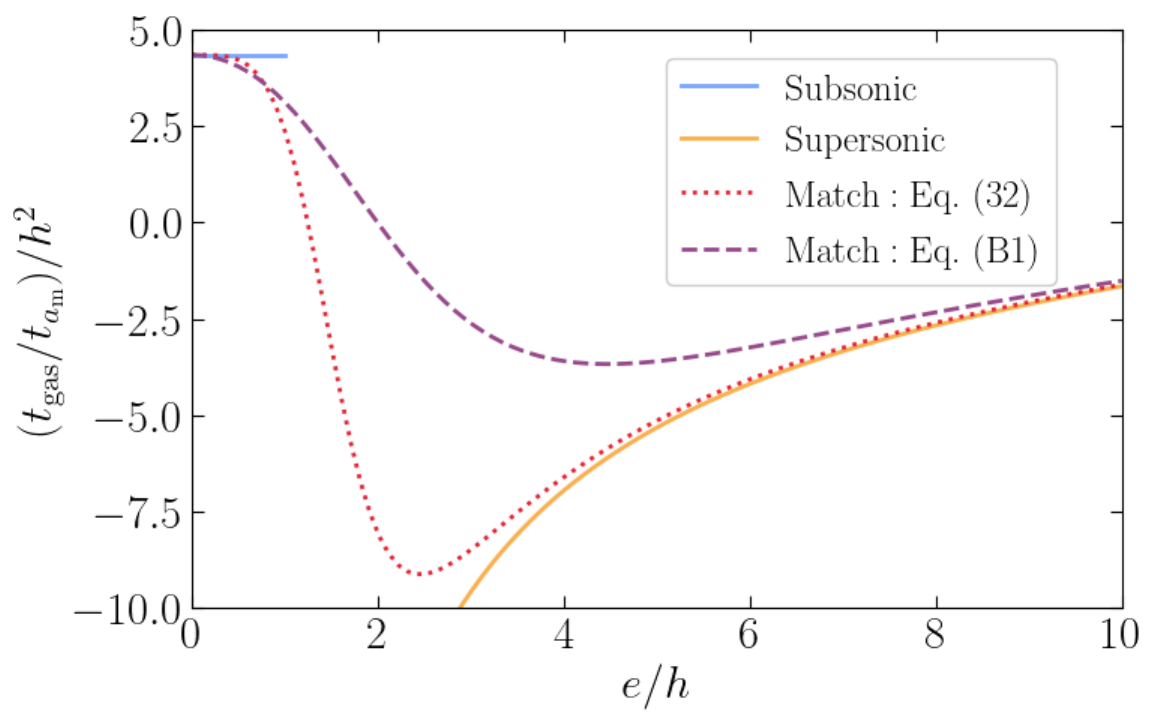}     
\caption{Comparison between two phenomenological models for the evolution timescale of the semi-major maxis $a_\text{m}$ due to the interaction with the gas disk. One contains a bump near the transonic region ($e \sim 2h$) due to the resonant excitation of sound waves and corresponds to Eq.~\eqref{eq:ta_phenom}, used for the study in the main body of this work. The other has a ``milder'' transition between the supersonic and subsonic motion and corresponds to Eq.~\eqref{eq:ta_nobump}.}
\label{fig:ta_comp}
\end{figure}
\begin{figure}[t]
\includegraphics[width=0.99\linewidth]{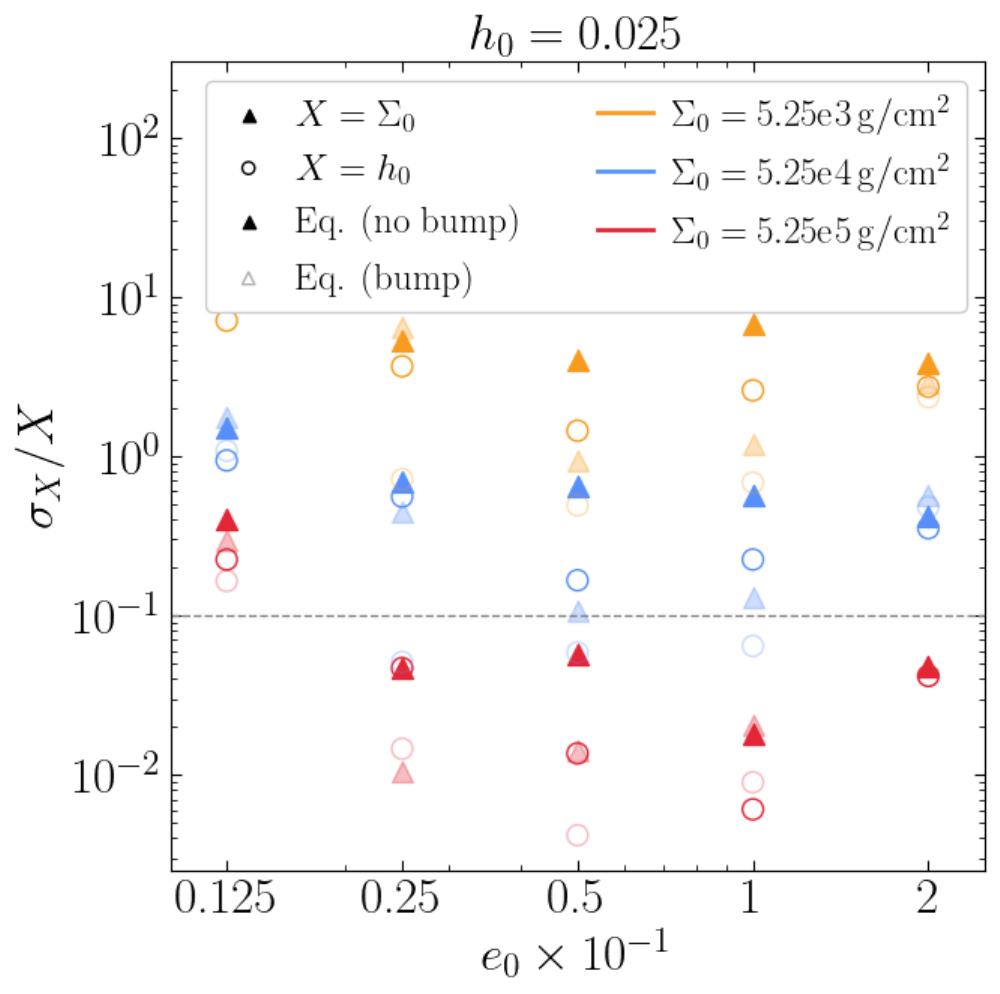}     
\caption{Same as in the left panel of Fig.~\ref{fig:SigmaNoConversion}, but for the ``milder'' model for the correction to the evolution of the semi-major axis $(a_\text{m})_\text{gas}$, given by Eq.~\eqref{eq:ta_nobump} in this Appendix and illustrated in Fig.~\ref{fig:ta_comp}. We also plot in faded colour the results shown in the left panel of Fig.~\ref{fig:SigmaNoConversion} to facilitate comparison between the two models.}
\label{fig:ConstraintsComp}
\end{figure}

In this Appendix, we explore the impact on our results from imposing a different matching condition between supersonic and subsonic motion for the evolution of the orbital parameters \textcolor{black}{due to the interaction with the gas in the inner disk region}. In particular, we prescribe a milder transition for the semi-major axis, removing the bump present at $e \sim 2h$ in $t_\text{gas}/t_{a_\text{m}}$~\eqref{eq:ta_phenom} (right panel of Fig.~\ref{fig:te_ta}). We reiterate, however, that an enhancement of the disk interaction is expected near the transonic region due to formation of shock waves. Nonetheless, considering astrophysical uncertainties and discrepancies between different hydrodynamical codes, we seek to evaluate how significant the increase in $(\dot{a}_\text{m})_\text{gas}$ is to the constraints obtained for the disk properties (Sec.~\ref{sec:Constraints} and specifically Fig.~\ref{fig:SigmaNoConversion}). We propose a different phenomenological model given by
\begin{equation}
\frac{t_\text{gas}}{t_{a_\text{m}}} = 2 C_\text{sub} \, h^2 \left(1-e^2\right)\frac{1-\left(\frac{e}{2h}\right)^2}{1+\left(\frac{e}{3h}\right)^3} \, . 
\label{eq:ta_nobump}
\end{equation}
The comparison between this and the model used in Eq.~\eqref{eq:ta_phenom} is illustrated in Fig.~\ref{fig:ta_comp}.

We repeat the study done in Section~\ref{sec:Constraints} using the ``milder'' model, for the same EMRI system (full parameters listed in Table~\ref{tab:EMRIparameters}) and the different disk configurations. Our results for the measurement uncertainty for the central surface density and aspect ratio of the disk as a function of the initial eccentricity are shown in Fig.~\ref{fig:ConstraintsComp} for $h_0 = 0.025$ (which, for the ``stronger'' model is shown in the left panel of Fig.~\ref{fig:SigmaNoConversion}). We also plot in faded colours the constraints obtained with the main model to facilitate comparison between the two.

For most of the parameter space, the model without the enhancement in the transonic region leads to weaker constraints, which is intuitive considering the relative amplitude of the gas effect in the evolution of the orbital separation with respect to the GW radiation reaction is smaller. Nonetheless, the difference in the measurement uncertainty ($\sigma_X/X$) is almost always a factor of $\mathcal{O}(1)$, and does not have an impact in the "measurability criterion", i.e. the constraints for the different prescritpions do not cross the dashed horizontal line in Fig.~\ref{fig:ConstraintsComp}. In fact, for some disk configurations the constraints obtained with the model without the bump~\eqref{eq:ta_nobump} are better, highlighting the need for more detailed modelling of disk interactions in EMRI evolutions.

\section{Full corner}\label{app:CornerFisher}

\begin{figure*}[t]
\includegraphics[width=1\linewidth]{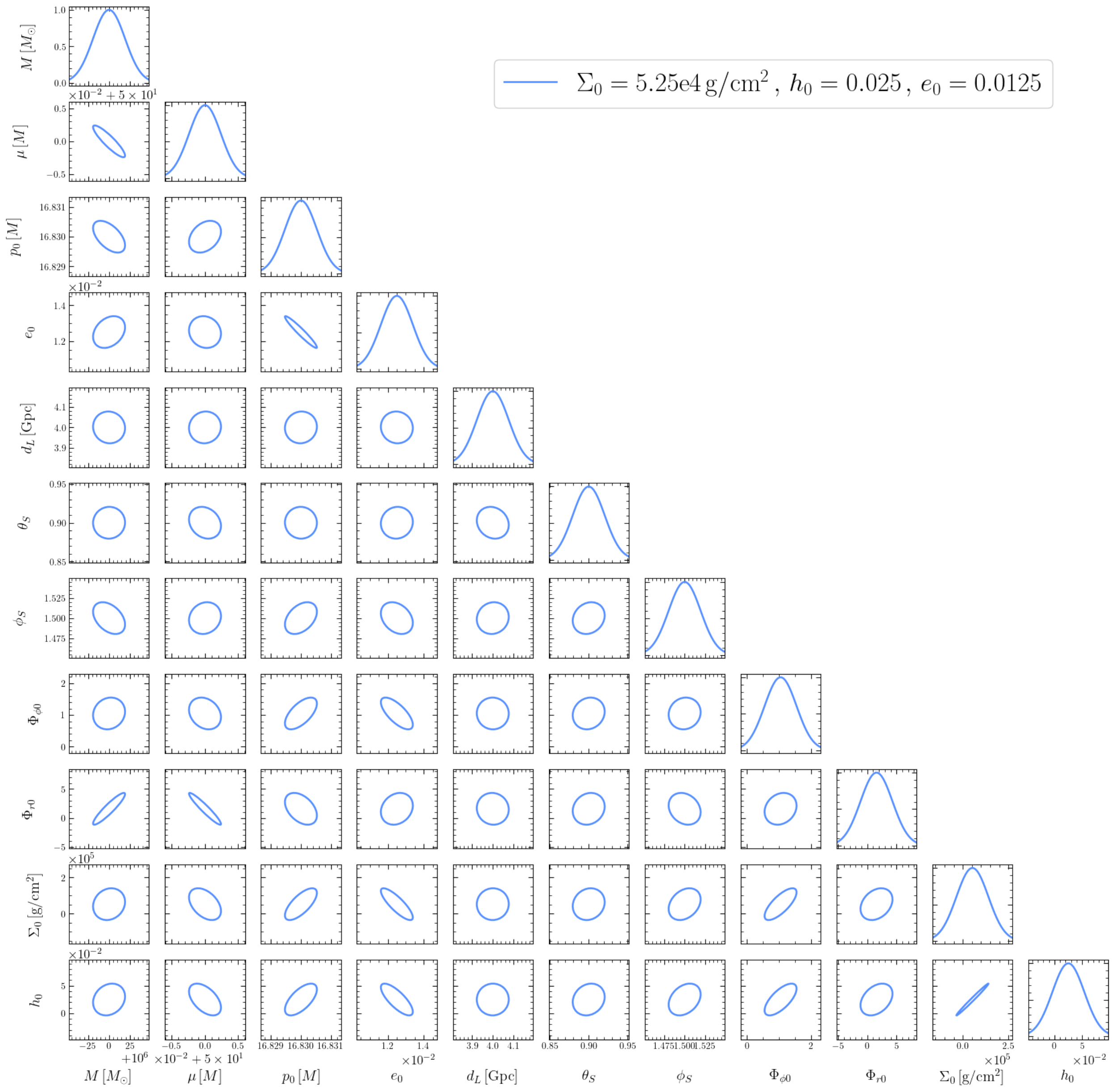}    
\caption{Full corner plot computed via the Fisher-Information Matrix for the EMRI system in Table~\ref{tab:EMRIparameters}, with initial eccentricity $e_0 = 0.0125$ and disk configuration $\Sigma_0 = 5.25\times 10^{4} \, \text{g}/\text{cm}^2$ and $h_0 = 0.025$. The EMRI is always in the subsonic regime and its eccentricity after 4 years of inspiral is $e_f = 3.3 \times 10^{-3}$. As mentioned in Table~\ref{tab:EMRIparameters}, quoted masses are in the detector frame. This corresponds to measurement precisions for the source frame masses of $\sigma_{M_S}/M_{S}, \sigma_{\mu_S}/\mu_{S} \sim 10^{ -2} $ (where $S$ denotes source frame).}
\label{fig:FullCorner_Subsonic}
\end{figure*}

\begin{figure*}[t]
\includegraphics[width=1\linewidth]{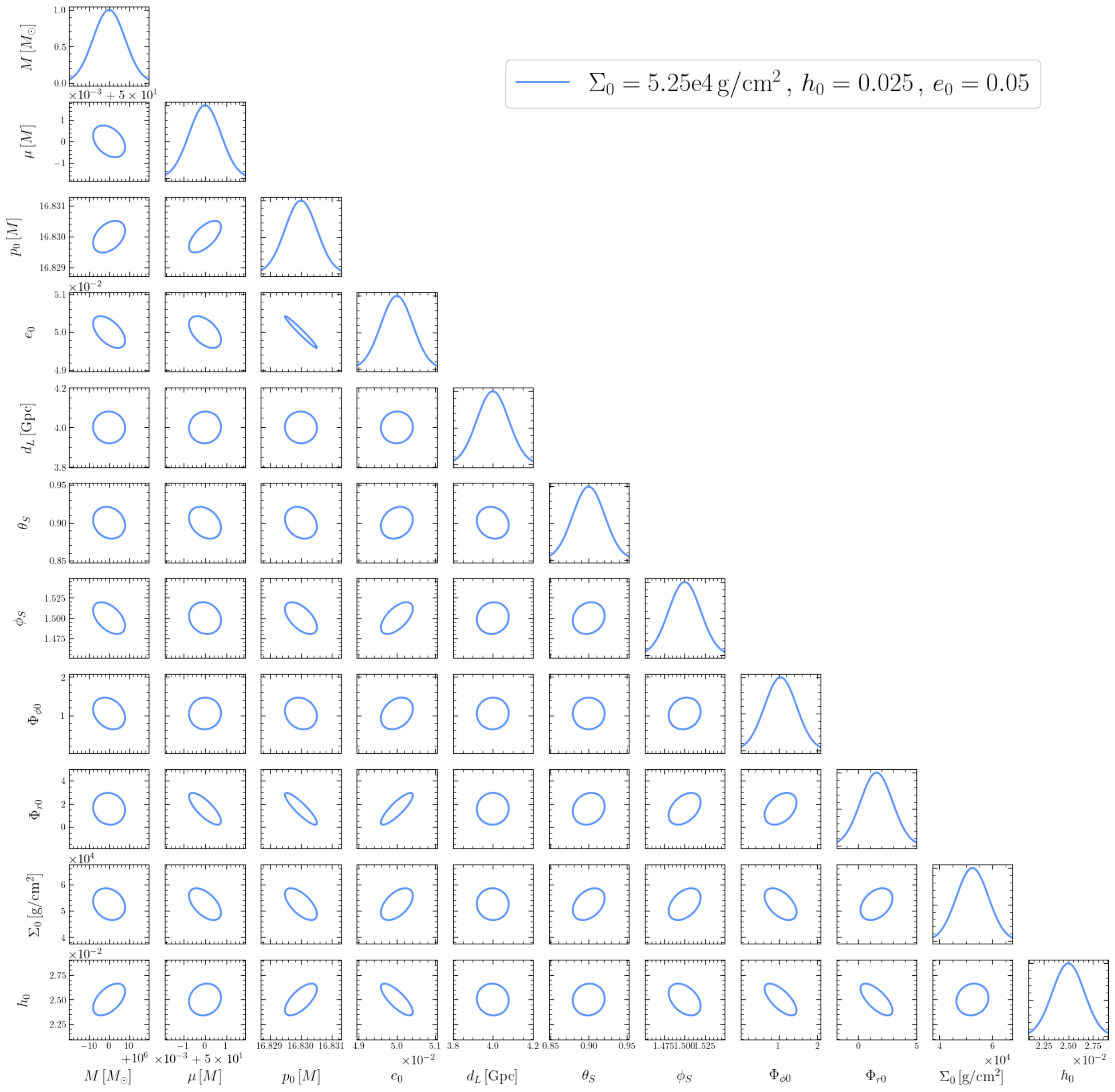}    
\caption{Same as in Fig.~\ref{fig:FullCorner_Subsonic} for initial eccentricity $e_0 = 0.05$. The EMRI starts in supersonic motion and transitions to subsonic one during the inspiral, having  $e_f = 1.3\times 10^{-2}$ after 4 years of evolution. Notice that the $M-p_0$, $M-e_0$, $p_0-\Sigma_0$ correlations change sign with respect to the fully subsonic case. Parameters have the same interpretation as in Table~\ref{tab:EMRIparameters} and Fig.~\ref{fig:FullCorner_Subsonic}. }
\label{fig:FullCorner_Supersonic}
\end{figure*}

In this Appendix, we show two examples of the full corner plot computed via the Fisher Information Matrix for the same EMRI system (Table~\ref{tab:EMRIparameters}) and disk configuration ($\Sigma_0 = 5.25\times 10^{4} \, \text{g}/\text{cm}^2$, $h_0 = 0.025$) and two different initial eccentricities. One corresponds to a fully subsonic inspiral ($e_0 = 0.0125$ in Fig.~\ref{fig:FullCorner_Subsonic}) and another to an EMRI that starts in the supersonic regime and transitions to subsonic during the inspiral ($e_0 = 0.05$ in Fig.~\ref{fig:FullCorner_Supersonic}).


\end{document}